\documentclass[12pt,epsf,amssymb]{article}

\usepackage{graphicx}
\usepackage{amsfonts}
\usepackage{amsmath}
\usepackage[usenames]{color}
\usepackage{amssymb}
\usepackage[mathscr]{eucal} 
\usepackage{url}

\def\Z{\mathbb{Z}}
\def\Q{\mathbb{Q}}
\def\R{\mathbb{R}}
\def\C{\mathbb{C}}
\def\P{\mathbb{P}}

\begin{document}

\baselineskip 0.6cm
\newcommand{\vev}[1]{ \left\langle {#1} \right\rangle }
\newcommand{\bra}[1]{ \langle {#1} | }
\newcommand{\ket}[1]{ | {#1} \rangle }
\newcommand{\Dsl}{\mbox{\ooalign{\hfil/\hfil\crcr$D$}}}
\newcommand{\nequiv}{\mbox{\ooalign{\hfil/\hfil\crcr$\equiv$}}}
\newcommand{\nsupset}{\mbox{\ooalign{\hfil/\hfil\crcr$\supset$}}}
\newcommand{\nni}{\mbox{\ooalign{\hfil/\hfil\crcr$\ni$}}}
\newcommand{\nin}{\mbox{\ooalign{\hfil/\hfil\crcr$\in$}}}
\newcommand{\Slash}[1]{{\ooalign{\hfil/\hfil\crcr$#1$}}}
\newcommand{\EV}{ {\rm eV} }
\newcommand{\KEV}{ {\rm keV} }
\newcommand{\MEV}{ {\rm MeV} }
\newcommand{\GEV}{ {\rm GeV} }
\newcommand{\TEV}{ {\rm TeV} }

\def\diag{\mathop{\rm diag}\nolimits}
\def\tr{\mathop{\rm tr}}

\def\Spin{\mathop{\rm Spin}}
\def\SO{\mathop{\rm SO}}
\def\O{\mathop{\rm O}}
\def\SU{\mathop{\rm SU}}
\def\U{\mathop{\rm U}}
\def\Sp{\mathop{\rm Sp}}
\def\SL{\mathop{\rm SL}}

\def\change#1#2{{\color{blue}#1}{\color{red} [#2]}\color{black}\hbox{}}

\begin{titlepage}
  
\begin{flushright}
IPMU15-0097
\end{flushright}
 
 \vskip 1cm
 \begin{center}
 
 {\large \bf Statistics of Flux Vacua for Particle Physics}
 
 \vskip 1.2cm
 

Taizan Watari
 
\vskip 0.4cm
%
%
   Kavli Institute for the Physics and Mathematics of the Universe, 
   University of Tokyo, Kashiwa-no-ha 5-1-5, 277-8583, Japan

\vskip 1.5cm
   
\abstract{Supersymmetric flux compactification of F-theory in the geometric phase yields numerous 
vacua, and provides an ensemble of low-energy effective theories with different symmetry, matter 
multiplicity and Lagrangian parameters. Theoretical tools have already been developed so that we 
can study how the statistics of flux vacua depend on the choice of symmetry and some of Lagrangian 
parameters. 
In this article, we estimate the fraction of i) vacua that have a U(1) symmetry for spontaneous 
R-parity violation, and ii) those that realise ideas which achieve hierarchical eigenvalues of the 
Yukawa matrices. We also learn a lesson that the number of flux vacua is reduced very much when 
the unbroken $\U(1)_Y$ symmetry is obtained from a non-trivial Mordell--Weil group, while it is not, 
when $\U(1)_Y$ is in SU(5) unification. It also turns out that vacua with an approximate U(1) 
symmetry forms a locus of accumulation points of the flux vacua distribution.  } 
  
\end{center}
\end{titlepage}
 

\section{Introduction}

Flux compactification of F-theory/Type IIB string theory generates discretum of vacua in the complex 
structure parameter space, making it possible to {\it count} vacua and argue statistics of some of 
observables in the low-energy effective theories \cite{flux-cpt-review, Denef}. It is virtually 
impossible to work out the vacuum for each one of individual flux configurations in practice, but 
this difficulty can be overcome in an approximate treatment of this problem introduced by 
Ashok--Denef--Douglas \cite{ADD-1, ADD-2}. Their treatment becomes a very powerful tool, when used 
for F-theory compactifications \cite{Denef, 1401Yusuke}, since one can estimate the number of flux 
vacua that lead to low-energy effective theories with a given set of 7-brane gauge groups and the 
number of generations of matter fields. It turns out \cite{1408Andreas} that the number of 
flux vacua is reduced in the order of $10^{-O(100)}$ generically as we require the rank of 7-brane 
gauge group to be higher by one. Focusing on an ensemble of flux vacua with a given 7-brane gauge 
group, one further finds that the number of flux vacua follows the Gaussian 
distribution on the number of generations $N_{\rm gen}$, with the variance $\vev{N_{\rm gen}^2}$ 
not more than O(1).

Obviously the analysis method above can be applied also to more refined and practical problems. 
It often happens in model building that more than one theoretically and phenomenologically 
consistent idea (model) has been proposed for a given phenomenon, and one cannot say which is 
better within the framework of low-energy effective field theory. By counting the number of 
flux vacua that realise various ideas and comparing the numbers, however, one can introduce 
a measure of naturalness on those consistent ideas. Such attempts have been made often in Type IIB
compactifications so far; we are returning to this program by using F-theory compactifications so 
that we can address questions involving non-Abelian/Abelian gauge groups in the low-energy 
effective theories.  

There are two kinds of naturalness/statistics questions. Note first that a low-energy effective 
theory is specified by providing a set of model data; a set of data consists of algebraic 
data (e.g. symmetry), topological data (e.g., matter multiplicity) and moduli data (i.e. 
coupling constants, symmetry breaking scale, etc.). Since a choice of algebraic and topological 
data is discrete in nature, we ask such questions as how much fraction of flux vacua survives 
when a certain symmetry is imposed. Section \ref{sec:fraction} is devoted to this category of 
problems. Moduli data, on the other hand, show up as continuous parameters in effective theories, 
and the flux vacua statistics need to be presented as a continuous distribution on the parameter 
space. This question is addressed by using F-theory compactification in 
section \ref{sec:distribution}. 

Section \ref{sec:fraction} deals with 
\begin{itemize}
\item dimension-4 proton decay: spontaneous R-parity violation (v.s. $\Z_2$ symmetry),
\item SU(5) unification v.s. $\SU(3) \times \SU(2) \times \U(1)$ without unification.
\end{itemize}
We do not get our hands on discrete symmetry in this article; we just estimate statistical cost 
of introducing an extra U(1) symmetry, which is relevant to both of the physics questions 
above. Section \ref{sec:distribution} begins with a recap of \cite{taxonomy, ET}; observations 
made in these articles---originally in Type IIB context---holds readily in F-theory 
compactifications. We then discuss 
\begin{itemize}
\item distribution of symmetry breaking scale of an approximate U(1) symmetry, 
\item two solutions to the hierarchical structure problem of Yukawa matrices.
\end{itemize}
The first and last of the four subjects above are found in the list of possible applications 
in \cite{1408Andreas}.
The appendix \ref{sec:review-MW} is a brief review note on two constructions of fourfold geometry 
for F-theory compactifications with a U(1) symmetry; the appendix \ref{sec:SU(6)} provides a little 
more details about SU(6) unification models with up-type Yukawa coupling in F-theory than in the 
literature. Monodromy of four-cycles in a fourfold is studied in the 
appendix \ref{ssec:cycle-monodromy}, as we need the result in section \ref{ssec:approx-U1-general}.

\section{A Quick Review of the Formulation}
\label{sec:ADD}

Suppose that one is interested in estimating the number of flux vacua which have a given set of 
algebraic and topological properties in the effective theory below the Kaluza--Klein scale. 
Once we specify topology of the base threefold $B_3$ and of the divisor class $S \subset B_3$ 
supporting unification gauge group (7-brane),\footnote{In this article, except in 
section \ref{ssec:GUTvsMSSM},  we use the word unification group and non-Abelian 7-brane gauge 
group interchangeably, because gauge coupling unification is guaranteed when a flux on $S$ 
breaks the non-Abelian gauge group symmetry on $S$ to its subgroup $G_1 \times G_2 \times \cdots$.} 
we can think of a family of non-singular Calabi--Yau fourfolds $\hat{Y}_4$ with elliptic fibration 
over $B_3$ consistent with the set of algebraic properties one is interested in. Let ${\cal M}_*$ 
be the space of complex structure parameters for this family.\footnote{We avoid using the term 
``moduli space'' for this meaning for the most part in this article. The space ${\cal M}_*$
is introduced and used in the present context just as a mathematical construct on which the 
result (vacuum index density $d\mu_I$) is presented, not as the non-linear sigma model target 
space in some approximation scheme of low-energy effective theory; once flux is introduced, these 
two notions are not the same. We hope to make this distinction clear 
by avoiding the word ``moduli space'' for the former, although it is perfectly 
correct to refer to the former as a moduli space in math context. } 
Statistics of flux vacua should turn out as a scatter plot on this parameter space ${\cal M}_*$. 
When the ensemble of topological flux configurations is replaced by its continuous 
approximation \cite{ADD-1, ADD-2}, the scatter plot of vacua turns into a vacuum distribution 
function (an $(m,m)$-form on ${\cal M}_*$; $m := {\rm dim}_\C {\cal M}_*$). 
Ashok--Douglas \cite{ADD-1} introduced vacuum index density $d\mu_I$, to which individual flux 
vacua contribute by $\pm 1$ (rather than by $+1$). It is also an $(m,m)$-form on ${\cal M}_*$ 
under the continuous approximation, and is much easier to compute \cite{ADD-1, ADD-2}. For this 
practical reason, we also use the vacuum index density $d\mu_I$ in this article, instead of the 
vacuum density. 

The vacuum index density turns out to have the following expression 
\cite{ADD-1, ADD-2, Denef, 1401Yusuke}:\footnote{The prefactor for $L_* \ll K$ was discussed 
in \cite{ADD-1}, but was corrected in \cite{Denef, 1408Andreas}.}
\begin{equation}
  d\mu_I \sim
 \left\{ \begin{array}{ll} 
     \frac{(2\pi L_*)^{K/2}}{(K/2)!} & {\rm if~} K \ll L_* \\
     \frac{K^{L_*}}{L_*!} & {\rm if~} L_* \ll K 
  \end{array} \right\} 
 \times \rho_I, \qquad 
  \rho_I = {\rm det}_{m \times m}
     \left( - \frac{R}{2\pi i} + \frac{\omega}{2\pi} {\bf 1}_{m \times m} \right).
\label{eq:ADD-formula}
\end{equation}
Here, $R$ is the curvature two-form of $T{\cal M}_*$ and $\omega$ the K\"{a}hler form on 
${\cal M}_*$. $K$ is the dimension of an Affine subspace  
\begin{equation}
  \left\{ G_{\rm fix} + \Delta G \; | \; \Delta G \in H_{\rm scan} \right\} \subset H^4(\hat{Y}_4; \R)
\end{equation}
in which the four-form flux is scanned freely; $H_{\rm scan}$ is a vector subspace of 
$H^4(\hat{Y}_4; \R)$, and $K := {\rm dim}_\R H_{\rm scan}$. $L_*$ is the upper bound on the 3-brane 
charge that the scanning component of the four-form flux $\Delta G$ contributes to. 
See \cite{1408Andreas} for more detailed explanations. For the ensemble of fluxes above to 
correspond to an inclusive enough ensemble of effective theories with a given set of algebraic and 
topological data, $H_{\rm scan}$ need to contain the primary horizontal component 
\begin{equation}
  \left[ H^{4,0}(\hat{Y}_4; \C) + {\rm h.c.} \right] \oplus 
  \left[ H^{3,1}(\hat{Y}_4; \C) + {\rm h.c.} \right] \oplus 
  H^{2,2}_H(\hat{Y}_4; \R).
\end{equation}
This condition on the minimum inclusiveness of flux ensemble is also known to be a necessary and 
sufficient condition for the formula of $\rho_I$ in (\ref{eq:ADD-formula}) to hold in F-theory 
compactification \cite{Denef, 1401Yusuke}. This means that 
\begin{equation}
  K \geq K_0 := 2(1 + h^{3,1}) + h^{2,2}_H.
\end{equation}
Specific physics questions of one's interest determine how inclusive an ensemble of flux vacua 
one wants to pay attention to, and how large a subspace of 
$H^{2,2}_V(\hat{Y}_4;\R) \oplus H^{2,2}_{RM}(\hat{Y}_4; \R)$ should be included in $H_{\rm scan}$; 
the choice of $K-K_0$ is discussed in an application to the spontaneous R-parity violation 
scenario in section \ref{ssec:R-parity-violation}; see also \cite{1401Yusuke, 1408Andreas}.

As the integral $\int \rho_I$ over a fundamental domain of ${\cal M}_*$ usually turns out to 
be a value of order unity, we can just use the prefactor in (\ref{eq:ADD-formula}) as an estimate 
of the number of flux vacua that have a set of algebraic and topological data specified at the 
beginning; we just use this prefactor for the study in section \ref{sec:fraction}. 
The distribution $\rho_I$ can be used to study statistical distribution of coupling constants / 
Lagrangian parameters within a class of low-energy effective theories with a given set of algebraic 
and topological properties; this $\rho_I$ is used for the study in section \ref{sec:distribution}.

One needs to keep in mind that the distribution as well as the estimate of the number of flux vacua 
here does not require that the vacuum expectation value (vev) of superpotential is much smaller 
than the Planck-sclae-cubed; large fraction of vacua have AdS supersymmetry. 
Stabilisation of K\"{a}hler moduli is not studied either. For these reasons and for other reasons 
stated elsewhere in this article, the formula (\ref{eq:ADD-formula}) should be regarded only 
as partial information of statistical distribution of observables in string landscapes. 

\section{Fraction of Flux Vacua with Enhanced Symmetries}
\label{sec:fraction}

\subsection{Statistical Cost of Spontaneous R-parity Violation}
\label{ssec:R-parity-violation}

Dimension-4 proton decay problem in supersymmetric Standard Models can be avoided, for example, 
by either imposing a $\Z_2$-symmetry (matter/R-parity) or assuming spontaneous breakdown of 
a U(1) symmetry triggered by a non-zero Fayet--Iliopoulos parameter (spontaneous R-parity 
violation).\footnote{For right-handed neutrinos to be able to have large Majorana masses, 
it is better that the U(1) symmetry is broken at high energy. Despite the spontaneous 
breaking, the SUSY-zero mechanism \cite{susy0} remain at work in getting rid of dangerous 
proton decay operators at least for some UV constructions (see \cite{Kuriyama} for discussion).}
When we assume that the $\Z_2$ symmetry originates from a $\Z_2$ symmetry of a geometry for 
compactification, complex structure parameters of the geometry need to be in a special sub-locus 
for enhancement of the $\Z_2$ symmetry \cite{Het-Z2-tune}, and the flux vacua that end up in 
such a sub-locus will constitute small fraction of all the flux vacua \cite{Dine} (see also 
a remark at the end of this section \ref{ssec:R-parity-violation}). The spontaneous R-parity 
violation scenario (see \cite{TW-06, Kuriyama, TTW-RHnu} for its string implementation) also 
requires tuning, because we need a U(1) symmetry. This tuning should be translated into 
restriction on flux configuration. In this section \ref{ssec:R-parity-violation}, we estimate 
the fraction of flux configurations that have an extra U(1) symmetry. Comparing the faction of 
flux vacua for the spontaneous R-parity violation and that for matter/R-parity, one could 
argue which is solution to the dimension-4 proton decay problem is more ``natural'' in terms of 
flux vacua statistics. 

There are two different ways to implement an extra U(1) symmetry in F-theory 
compactifications.\footnote{There may be more, in fact, as we discuss in 
section \ref{ssec:GUTvsMSSM}.} 
One is to assume a 7-brane locus $S \times \R^{3,1}$ with an SU(6) or SO(10) gauge group, and 
introduce a U(1) flux on the complex surface $S$, so that the symmetry is broken\footnote{The 
F-theory implementation of spontaneous R-parity violation scenario is always an example of 
``T-brane'' \cite{Tbrane}. The D-term condition $\sum_i q_i |\phi_i|^2 - \xi = 0$ in the 4D effective 
theory corresponds \cite{DW-1, BHV-1} to a (D-term) BPS condition 
$[\varphi, \overline{\varphi}] + \omega \wedge F=0$ in the effective field theory on 
$S \times \R^{3,1}$ (Katz--Vafa type field theory \cite{KV}). The off-diagonal components of 
the Higgs field vev $\vev{\varphi}$ is therefore essential in the spontaneous R-parity violation 
scenario \cite{TW-06, TTW-RHnu}.} from SU(6) or SO(10) to 
$\SU(3)_C \times \SU(2)_L \times \U(1)_Y \times \U(1)$ \cite{TW-06, TTW-RHnu}. 
The other \cite{Caltech-xxxx, Hayashi-More, GW} is to get an extra U(1) symmetry by assuming 
a Calabi--Yau fourfold with a non-trivial Mordell--Weil group \cite{MV-2}. In the latter 
implementation, more variety is available in the choice of U(1) charge assignment than those 
that follow from Heterotic string geometric (supergravity) compactification \cite{VB-G-K, VB-G-K2, 
MW-U1-variety}.

\begin{center}
.............................................................
\end{center}

To get started, let us first take a moment to consider how one should choose $H_{\rm scan}$ for 
this problem. We address this question by working on a few concrete examples. First of all, 
the base threefold is set to be $B_3 = \P^1 \times \P^2$, and we require $\SU(5)$ 7-branes along 
a divisor $S = H_{\P^1} = {\rm pt} \times \P^2$ in $B_3$. There is a wide variety in constructing 
families of Calabi--Yau fourfolds with a non-trivial Mordell--Weil group\footnote{We do not work on 
the $H_{\rm scan}$ determination problem for the SU(6) or SO(10) realisation of spontaneous 
R-parity violation in this article. That will be a doable problem. As we see later, however, 
precise determination of $H_{\rm scan}$ is not much of importance 
when $h^{3,1} \gg h^{1,1}$.} \cite{MW-U1-variety}, but we just pick up only two of them to 
work on; in both of the two constructions, a Calabi--Yau fourfold $Y_4$ is obtained as a 
hypersurface of an ambient space that has a toric surface fibration over the base manifold $B_3$; 
the fibre surface is a blow-up of $WP^2_{[1:2:3]}$ in one of the two, and it is $F_1 = dP_1$ in the 
other. The appendix \ref{sec:review-MW} provides a brief summary note of the facts about the 
two constructions.

In the first construction (see the appendix \ref{sssec:BlWP123}), where 
${\rm Bl}_{[1:0:0]}WP^2_{[1:2:3]}$ is the fibre of the ambient space, the vertical component of 
$H^{2,2}$, namely, $H^{2,2}_V(\hat{Y}_4; \Q)$, is of 11 dimensions. Four among them are generated by 
\begin{equation}
 \sigma_0 \cdot H_{\P^1}, \qquad 
 \sigma_0 \cdot H_{\P^2}, \qquad 
 H_{\P^1} \cdot H_{\P^2}, \qquad 
 H_{\P^2} \cdot H_{\P^2}, 
\label{eq:vertical-1}
\end{equation}
where $\sigma_0$ is a zero section of $\pi: \hat{Y}_4 \longrightarrow B_3$ and $H_{\P^2}$ the 
hyperplane divisor of $\P^2$. Four other generators are the vanishing two-cycles of rank-4 SU(5) 
symmetry fibred over $H_{\P^2}|_S$: 
\begin{equation}
  E_{a} \cdot H_{\P^2} \qquad \qquad (a=1,2,3,4),
\label{eq:vertical-2}
\end{equation}
where $E_a$'s are the Cartan divisors of SU(5). All the three remaining generators are vanishing 
cycles associated with charged matter fields; two are for $\bar{\bf 5}_{-2}$ and $\bar{\bf 5}_{+3}$ 
representations of the $\SU(5) \times \U(1)$ symmetry, and the last one for the ${\bf 1}_{5}$ 
representation. 
The dimension of the remaining (i.e., non-horizontal non-vertical) component is determined by 
using the formula of \cite{1408Andreas}; it turns out that $h^{2,2}_{RM} = 0$. 

How should we choose $H_{\rm scan}$, then? 
First of all, the four-form $\Delta G$ needs to stay away from the 8 four-cycles listed 
in (\ref{eq:vertical-1}, \ref{eq:vertical-2}) in order not to break SO(3,1) and SU(5) unification 
symmetry.\footnote{We ignore SU(5) symmetry breaking to the Standard Model gauge group in this 
article, in order not to be distracted by unessential details.} Secondly, the net chirality 
``$N_{\rm gen}$'' of $\bar{\bf 5}_{-2}$ and $\bar{\bf 5}_{+3}$ need to be fixed, which means 
that the integral of a four-form over these two cycles need to have values designated by a 
phenomenology (low-energy) model of interest. Therefore, there should not be scanning of 
$\Delta G$ in the 8+2 dimensions of $H^{2,2}_{RM}(\hat{Y};\R) \oplus H^{2,2}_V(\hat{Y}_4; \R)$.
The net chirality of the ${\bf 1}_5$ field, however, may be chosen arbitrarily, as they do not 
appear in the low-energy spectrum in the spontaneous R-parity violation scenario.\footnote{This 
argument is a little simplified too much for phenomenology, but we keep the story simple in this 
article (see \cite{Kuriyama} for more). After all, small changes in the argument in these 
paragraphs do not severely affect the qualitative conclusion we draw later in this section.} 
Thus, this means that the four-form flux quanta can be scanned also in a one dimensional subspace 
of $H^{2,2}_{RM}(\hat{Y};\R) \oplus H^{2,2}_V(\hat{Y}_4; \R)$ for the question we are facing. This 
brings us to 
\begin{equation}
 K = K_0 + 1.
\end{equation}

Let us also work on one more construction of non-trivial Mordell--Weil group, where the ambient 
space of $\hat{Y}_4$ has $F_1$ fibre (see a review in the appendix \ref{sssec:F1}). The construction 
comes with topological choice of two divisors $\kappa^1$ and $\kappa^2$ on $B_3$; we stick to 
the same choice of $(B_3, [S])$ as before for now. The choice of the divisor classes change 
the topological class of various matter curves, but U(1) charge assignment is not affected. 
When the two divisors are parameterised by 
\begin{equation}
  \kappa^1 = a_1 H_{\P^1} + a_2 H_{\P^2}, \qquad 
  \kappa^2 = b_1 H_{\P^1} + b_2 H_{\P^2},
\end{equation}
we focus our attention to 
\begin{equation}
  b_1 = 0, \quad a_1 = 1,2, \quad 0 \leq a_2,b_2, \quad 
  0 \leq 6 +a_2-2b_2, \quad 
  0 \leq 6 -2a_2 + b_2, 
\end{equation}
since some of the coefficients $A_{0,1}$, $A_{1,0}$, $B_{-1,1}$, $B_{0,0}$, $b_{1,-1|2}$, $C_{-2,1}$, 
$c_{-1,0|1}$, $c_{0,1|3}$ and $c_{1,-2|5}$ in the defining equation of $\hat{Y}_4$ 
(\ref{eq:defeq-F1-fibre}, \ref{eq:def-ABC-F1fibre-gen}, \ref{eq:def-ABC-F1fibre-A4}) would vanish 
identically otherwise.\footnote{This constraint is not from physical reasons; when one of those 
constraints is not satisfied, it often happens that analysis of geometry is done better by using 
an ambient space that has a fibre other than $F_1 = dP_1$.} We further focus on cases with $a_2=0$, 
when the non-singular fourfold $\hat{Y}_4$ remains a flat fibration over $B_3$, and the low-energy 
spectrum is guaranteed to be free from tensionless string (cf \cite{tensionless}). This means that 
$0 \leq b_2 \leq 3$. 

We studied geometry associated with $H^{2,2}(\hat{Y}_4)$ carefully for $a_1 = 1$ and $0 < b_2 < 3$. 
The non-vertical and non-horizontal component $H^{2,2}_{RM}(\hat{Y}_4)$ turns out to be trivial, which 
follows from the formula in \cite{1408Andreas}. The vertical component $H^{2,2}_V(\hat{Y}_4;\R)$ has 
13 independent generators. The five independent generators other than those 
in (\ref{eq:vertical-1}, \ref{eq:vertical-2}) all correspond to the vanishing cycles associated 
with charged matter fields. Three correspond to $\bar{\bf 5}_0$, $\bar{\bf 5}_1$ and 
$\bar{\bf 5}_{-1}$, and two others to ${\bf 1}_1$ and ${\bf 1}_2$. 
Repeating the same argument as in the case of the first construction, we find that $H_{\rm scan}$
has a dimension 
\begin{equation}
  K = K_0 + 2.
\end{equation}

Spontaneous R-parity violation is a little special in that the U(1) symmetry exerts some 
controlling power on types of interactions in the low-energy effective theory even after it 
is broken spontaneously at high-energy (primarily for dimension-4 operators, not necessarily on 
non-renormalisable operators; see \cite{Kuriyama} for discussion). Chirality is not well-defined 
any more, however, for SU(5)-neutral U(1)-charged matter fields after the spontaneous breaking 
of the U(1). Without the chirality protection, they do not survive in the low-energy 
spectrum.\footnote{By ``low-energy'' and ``high-energy'', we mean ${\cal O}(1\mbox{--}1000)$ TeV 
and ${\cal O}(10^{13\mbox{--}16})$ GeV in this paragraph, whereas we also use the term low-energy 
(effective theory) in the sense that an intended energy scale is below the Kaluza--Klein scale. 
It will not be difficult to figure out from the context in which meaning ``low-energy'' is used.}
For this reason, when we count the number of flux vacua that realise spontaneous R-parity violation 
scenario, it is appropriate that the flux quanta changing the net chirality of SU(5)-neutral 
U(1)-charged fields should be scanned, as we have discussed above in detail. Some part of the 
vertical component of $H^{2,2}(\hat{Y}_4)$ therefore contributes to the dimension $K$ of the 
scanning space of flux $H_{\rm scan}$, and $K > K_0$. 

\begin{center}
.............................................................
\end{center}

Let us now study the statistical cost of an extra U(1) symmetry. An easiest way to do that is 
to compute $L_*$ and $K$ for some concrete choices of $(B_3, [S])$, and work out the prefactor 
of (\ref{eq:ADD-formula}). Comparing the prefactor for the case with an SU(5)$\times$U(1) symmetry 
with the one for the case with just SU(5) unification, we can estimate the tuning cost of 
the spontaneous R-parity violation scenario. We will take this experimental approach first, 
by using $B_3 = \P^1 \times \P^2$ and $S = H_{\P^1}$ as before, and then discuss later how the 
tuning cost depends on the choice of $(B_3, [S])$.

It takes extra efforts to compute the dimension of the horizontal component $h^{2,2}_H$ (by 
using the formula in \cite{1408Andreas}) and $\chi(\hat{Y}_4)$ (which are used for $L_*$ 
in (\ref{eq:ADD-formula})), but there is a short-cut for such choices as $B_3 = \P^1 \times \P^2$.
So long as $h^{3,1}(\hat{Y}_4) \gg h^{1,1}(\hat{Y}_4)$ holds, which is the case for the topology of 
$(B_3, [S])$ we chose above, $H^{2,2}(\hat{Y}_4)$ is dominated by the horizontal component, 
i.e., 
\begin{equation}
  h^{2,2}(\hat{Y}_4) \sim h^{2,2}_H(\hat{Y}_4) \gg h^{2,2}_V(\hat{Y}_4),  \quad h^{2,2}_{RM}(\hat{Y}_4), 
\end{equation}
as experience in \cite{1408Andreas} shows. This is enough to see that \cite{Denef, 1408Andreas}
\begin{equation}
 L_* \sim \frac{\chi(\hat{Y}_4)}{24} \sim \frac{b_4}{24} \sim \frac{K}{24}, \qquad 
 \frac{K^{L_*}}{(L_*)!} \sim \exp \left[ \frac{b_4(\hat{Y}_4)}{24} \ln(24) \right].
\end{equation}
Furthermore, if one is interested only in the ratio of two prefactors $K^{L*}/(L_*)!$ (relative 
tuning cost) in such geometries with $h^{3,1} \gg h^{1,1}$, a relation \cite{h22formula}
\begin{equation}
  h^{2,2} = 4(h^{3,1} + h^{1,1}) + 44 -2 h^{2,1}
\end{equation}
implies that $\Delta h^{2,2} \sim 4 \Delta h^{3,1}$, and $\Delta b_4 \sim 6 \Delta h^{3,1}$. All these 
combined allows us to estimate the relative tuning cost by \cite{1408Andreas}
\begin{equation}
  \exp \left[ \frac{\ln(24)}{4} \times (\Delta h^{3,1}) \right]
\label{eq:estimate-1}
\end{equation}
Numerically,\footnote{\label{fn:ln-24}
This numerical value should not be taken at face value. The underlying 
cohomology lattice of the flux scanning space $H_{\rm scan}$ is not necessarily unimodular, whereas 
the derivation of the prefactor $K^{L_*}/L_*!$ is stated in \cite{Denef, 1408Andreas} in its simplest 
form, where the underlying lattice is unimodular. The relative tuning cost in (\ref{eq:estimate-1}) 
should be read only as $\exp[{\cal O}(1) \times (\Delta h^{3,1})]$.} $[\ln(24)]/4 \simeq 0.8$. 
The fraction of flux vacua with an enhanced symmetry is determined in this expression by the 
number of complex structure parameters to be tuned. 

Now we only need to compute $h^{3,1}$'s and compare. 
\begin{eqnarray}
  WP_{[1:2:3]}\mbox{-fibred, ~no~gauge~group} & & h^{3,1} = 3277, \\
  WP_{[1:2:3]}\mbox{-fibred, ~SU(5)~gauge~group} & & h^{3,1}=2148,       \label{eq:h31-SU5} 
\end{eqnarray}
which are the reference values of $h^{3,1}$ for $(B_3, [S])$ we have chosen. In the spontaneous 
R-parity violation scenario realised in a rank-5 unification,  
\begin{eqnarray}
  WP_{[1:2:3]}\mbox{-fibred, ~SO(10)~gauge~group} & & h^{3,1} =2138,   \label{eq:h31-SO10} \\
  WP_{[1:2:3]}\mbox{~fibred, ~SU(6)~gauge~group} & & h^{3,1} \sim 1900. \label{eq:h31-SU6} 
\end{eqnarray}
The values of $h^{3,1}$ are taken from \cite{1408Andreas} for SU(5) and SO(10), and the value for  
SU(6) is computed in the appendix \ref{sec:SU(6)}. Among the Mordell--Weil implementations of the 
extra U(1), we have also computed $h^{3,1}$ for the two constructions referred to earlier (and 
reviewed in the appendix \ref{sec:review-MW}):
\begin{eqnarray}
  {\rm Bl}_{[1:0:0]}WP^2_{[1:2:3]} \mbox{-fibred} & & h^{3,1}=932, \label{eq:h31-BlWP} \\
  F_1 \mbox{-fibred,~no.2} & & h^{3,1} = 7 (b_2)^2 - 12 b_2 + 372, 
     \qquad  (a_1 = 1, \quad 0 < b_2 < 3).  \label{eq:h31-F1}
\end{eqnarray}

It turns out, for the $(B_3, [S])$ we chose, that the cost of Mordell--Weil implementations 
of spontaneous R-parity violation comes at the order of $e^{-1000}$, relatively to generic SU(5) 
unification; the number of flux vacua is reduced that much by requiring an extra U(1) symmetry 
through existence of a non-trivial section. In the other group of implementations, namely rank-5 
unifications with U(1) flux, the cost comes out as something like $e^{-10}$ for SO(10) and $e^{-200}$ 
for SU(6). All these cost estimates have been read out by comparing $h^{3,1}$ 
in (\ref{eq:h31-SO10}--\ref{eq:h31-F1}) with that in (\ref{eq:h31-SU5}).

It is tempting to argue, based on the numerical experiment for a single choice of $(B_3, [S])$ 
though, that the Mordell--Weil implementations of an extra U(1) tend to be much more costly than 
those through unification with one rank higher.\footnote{It is naive just to compare the tuning 
cost of those different implementations. The SO(10) and SU(6) implementations predict 
SO(10)-like and SU(6)-like flavour structure automatically, and the tuning cost for appropriate 
flavour structure in SU(5) unification is not the same as those in SU(6) or SO(10) unification. 
This tuning cost for flavour is significant in SO(10) unification, because both quark doublets 
and lepton doublets live on the same matter curve. The tuning cost for flavour in SU(6) 
unification will vary for its particle identifications; see discussion in the 
appendix \ref{sec:SU(6)}.} Plausible explanation will be that the Mordell--Weil 
implementations require more parameters to be tuned, because existence of an extra section restrain 
geometry over the entire base manifold $B_3$; the implementations through rank-5 unification, 
on the other hand, require higher order of vanishing of some sections along a divisor in $B_3$, 
which is a condition only on semi-local geometry. It is desirable, however, that this argument 
is either confirmed (or refuted instead) by $h^{3,1}$ computation for other constructions of 
Calabi--Yau's with a non-trivial Mordell--Weil group, and for other choices of $(B_3, [S])$.

\begin{center}
.............................................................
\end{center}

Studies show \cite{HW-appear} that Calabi--Yau fourfolds eligible for supersymmetric 
compactification of F-theory are distributed almost evenly in the $h^{3,1} \gg h^{1,1}$ corner 
and $h^{3,1} \ll h^{1,1}$ corner of the $h^{3,1}$--$h^{1,1}$ plane; this fits very well with an  
observation that a morphism of elliptic fibration to some threefold is allowed for large 
fraction of Calabi--Yau fourfolds with various topology \cite{most-CY-elliptic}.
Such a choice as $B_3 = \P^1 \times \P^2$, which we used for the numerical experiment above, 
ends up in the corner of $h^{3,1} \gg h^{1,1}$, and hence the estimates of the fraction of flux 
vacua with an extra U(1) symmetry is hardly typical values for all the possible topological 
choices of $(B_3, [S])$. 

It is not hard to find out how things go in the $h^{3,1}$--$h^{1,1}$ plane for various choices of 
$(B_3,[S])$, if we maintain $K$ close to $K_0$. Along a $h^{3,1} + h^{1,1} = {\rm const}$ 
line in the $h^{3,1}$--$h^{1,1}$ plane, the Euler number $\chi \sim 2 (h^{3,1} + h^{1,1})+h^{2,2} \sim 
6 (h^{3,1} + h^{1,1})$ and the value of $L_* \sim \chi/24$ do not change much, but the value of 
$K_0 \sim 2[1+h^{3,1}] + h^{2,2}_H$ increases toward the $h^{3,1} \gg h^{1,1}$ corner. The prefactor 
in (\ref{eq:ADD-formula}) is an increasing function of $K$ for a given $L_*$, regardless of 
whether $L_* \ll K$ or $L_* \gg K$. The more Fano-like $B_3$ is, the {\it ample}r sections 
are available to $(-K_B)^{\otimes {\rm positive}}$, the larger $h^{3,1}$ is, and the larger the 
number of flux vacua is, after all. When a stack of SU(5) 7-branes is required along 
$S \subset B_3$, more sections (and hence $h^{3,1}$, and the flux vacua) are lost when 
$B_3$ is more Fano like; the loss is severer, if $c_1(N_{S|B_3})$ is ``positive''.  
The relative tuning cost is higher for Fano-like $B_3$, with positive $c_1(N_{S|B_3})$.
Experience in \cite{1408Andreas} shows that the number of remaining flux vacua (i.e., those 
with an SU(5) symmetry) tends to be larger in Fano-like $B_3$ and positive $c_1(N_{S|B_3})$, 
despite the severer tuning cost for SU(5) 7-branes. The same story will hold, even when 
$\SU(5) \times \U(1)$ symmetry is required instead. 

Let us note that the qualitative argument above is naive in various respects. First, we set 
$K = K_0$ above for simplicity, but there is a large room for $K - K_0$, when $B_3$ is far 
from being Fano, and $c_1(N_{S|B_3})$ far from being ``positive''. Such a set-up is possible in 
F-theory compactification \cite{non-Higgsable, HW-appear}, and it is known in such cases that 
there can be many other 7-branes with non-Abelian gauge groups, and $h^{1,1} \gg h^{3,1}$ for 
the fourfolds. It is then expected from experience in \cite{1408Andreas} that 
$h^{2,2}_V \gg h^{2,2}_H$. One then need to ask how much four-form flux can be introduced in the 
vertical component $H^{2,2}_V(\hat{Y}_4;\Q)$ without breaking SO(3,1) symmetry and supersymmetry 
(if one wishes); based on an answer to this technical question, one can then wonder how 
inclusive an ensemble of low-energy effective theory one is interested in, and how large 
$(K-K_0)$ is. 
Secondly, particle physics with SU(5) unification is not all we need in this universe. Some 
source of supersymmetry breaking needs to be present. While anti-D3 branes may be able to play 
some role, dynamical supersymmetry breaking in a non-Abelian gauge theory (e.g. the 3-2 model 
in \cite{DSB}) might also be at work. The tuning-cost-free non-Abelian gauge group in the 
non-Higgsable cluster may have something to do with dynamical supersymmetry breaking. 
Thirdly, the K\"{a}hler moduli need to be stabilised without a tachyon. U(1) fluxes change the 
effective number of K\"{a}hler moduli to be stabilised non-perturbatively, through the 
Fayet--Iliopoulos D-term potential (primitiveness condition of the flux). Finally, inflation or 
cosmological evolution in general may introduce some preference in the choice of $(B_3, [S])$.
All these issues are beyond the scope of this article.

\begin{center}
............................................................
\end{center}

This article does not try to estimate the fraction of flux vacua with an unbroken matter/R-parity 
symmetry. If one is to argue which one of R-parity and spontaneous R-parity violation is more 
natural solution to the dimension-4 proton decay problem in terms of flux vacua statistics, we 
also need an estimate for the R-parity scenario.
Although there are earlier works on this issue in the context of Type IIB orientifold 
compactifications (e.g. \cite{Dine}), further study in F-theory is desirable. 
It is worth reminding ourselves that the fact that $L_* \ll K$ in cases of $h^{3,1} \gg h^{1,1}$ 
may have an important implication to this issue. Continuous approximation to the space of fluxes 
in \cite{ADD-1, ADD-2} is fairly good when $K \ll L_*$; intuitively, as in \cite{BP}, that is 
when the radius-square ($L_*$) of a $K$-dimensional ``sphere''\footnote{In reality, the lattice  
$H^4(\hat{Y}; \Z)$ is not positive definite. It still seems, however, that this ``intuition'' 
holds at least to some extent, because the Bousso--Polchinski like prefactor 
of (\cite{ADD-formula}) was obtained in \cite{ADD-1, ADD-2} without assuming that the lattice 
is positive definite.} is much larger than the number of dimensions $K$. In the case with 
$L_* \ll K$ (which is the case at least when $h^{1,1} \ll h^{3,1}$), however, much larger fraction 
of flux configurations may end up with special points in the complex structure parameter space 
(sometimes with an accidental discrete symmetry) than expected in the continuous 
approximation \cite{ADD-2, Enumerate}. 

\subsection{GUT's and $\SU(3) \times \SU(2) \times \U(1)$ }
\label{ssec:GUTvsMSSM}

Pursuit of supersymmetric SU(5) unification is a primary motivation to study F-theory 
compactification. Doublet--triplet splitting problem motivates compactification in the 
geometric phase (supergravity regime), rather than stringy regime, because it is solved in 
a simple way by topology (hypercharge line bundle or Wilson line) on an internal 
space \cite{Witten-DT}; the up-type Yukawa coupling of the form 
${\bf 10}^{ij} {\bf 10}^{kl} {\bf 5}^m \epsilon_{ijklm}$ hints at algebra of the exceptional 
Lie group $E_{6,7,8}$ \cite{TW-06}. There is no direct experimental evidence (such as proton 
decay) so far for unification, however; certainly renormalisation group of the 
minimal supersymmetric Standard Model (MSSM) is consistent with gauge coupling unification, 
but we do not know for sure what the particle spectrum is like at energy scale higher than TeV.
If one does not take SU(5) unification seriously, then string vacua based on CFT's with a 
non-geometric target space are perfectly qualified; we do not have to require that $E_{6,7,8}$  
algebra be relevant for ``compactification'' either. 

With this perspective in mind, it makes sense to ask a question which is more popular in the 
ensemble of supersymmetric vacua of F-theory compactification in the geometric phase, SU(5) 
unification or MSSM without unification. If there are more MSSM vacua without unification 
than those with SU(5) unification within the landscape of F-theory, the MSSM vacua will 
surely outnumber those with unification in the entire string landscape, which includes string 
vacua based on non-geometric CFT's. Democracy, or simple majority rule, may not be the ultimate 
vacuum selection principle of string theory, but this question will still be of interest for 
those who are concerned about particle physics. 

It is necessary, before answering the question above, to think what unification means. The 
motivation of unified theory at the very beginning \cite{GG} was to explain quantisation of 
hypercharges. This charge quantisation is achieved in any realisation of the Standard Model in 
F-theory/Type IIB string theory, however. Even when the U(1) hypercharge is not embedded into 
a larger non-Abelian group, charges of $(p,q)$ strings (or M2-branes) are determined by 
algebraic topology, and the charges turn out to be quantised. Charge quantisation is therefore 
not a distinction criterion of, or motivation for, unification from the perspective of string 
theory. 

Let us list up a couple of criteria for unified theories:
\begin{itemize}
 \item $\SU(3)_C \times \SU(2)_L \times \U(1)_Y$ originates from a single stack of branes,
 \item all of $\SU(3)_C$, $\SU(2)_L$ and $\U(1)_Y$ are understood in a semi-simple brane configuration
\item gauge coupling unification is explained automatically,
\item matter fields in some of the five irreducible representations of the Standard Model, 
$({\bf 3}, {\bf 2})_{1/6}$, $(\bar{\bf 3}, {\bf 1})_{-2/3}$, $(\bar{\bf 3}, {\bf 1})_{+1/3}$, 
$({\bf 1}, {\bf 2})_{-1/2}$ and $({\bf 1}, {\bf 1})_{+1}$, are localised in the same locus 
in the internal space. 
\end{itemize}
SU(5) GUT models discussed in section \ref{ssec:R-parity-violation} satisfy all of those 
criteria. On the other hand, none of those criteria are satisfied, if $\SU(3)_C$ and $\SU(2)_L$ 
come from 7-branes on topologically different divisors $S_3$ and $S_2$, respectively, and 
$\U(1)_Y$ from a non-trivial section in the Mordell--Weil group. 

There will be another class of constructions of supersymmetric Standard Models in F-theory 
compactifications. In Calabi--Yau orientifold compactification of Type IIB string theory, 
configuration of six intersectiong D7-branes may give rise to a 
$\U(3)_C \times \U(2)_L \times \U(1)_Y$ gauge group, and its anomaly-free subgroup may be 
identical to the Standard Model gauge group. One should keep in mind here that the 7-brane 
tadpole cancellation condition requires more D7-branes and O7-planes than the minimal set of 
six D7-branes for $\U(3) \times \U(2) \times \U(1)$. In F-theory language,\footnote{The author 
has not understood a systematic way to lift Type IIB Calabi--Yau orientifold with intersecting 
7-branes into F-theory language. This paragraph is written by assuming that U(1)'s on 
intersecting D7-branes do not necessarily correspond to Mordell--Weil U(1)'s in F-theory lift 
(when the lift exists); if this assumption is wrong, this paragraph should be simply ignored. } 
the combination of four $A$-branes and a pair of $B$ and $C$-branes needs to be used as a 
package (while allowing deformation and intersection), and the elliptic fibration should look 
like the Seiberg--Witten geometry for $\SU(N_c = 2)$ $N_f = 4$ gauge theory (with mass 
perturbation) glued together, for any curve in $B_3$. In Type IIB language, it seems as if 
an extra U(1) symmetry is always obtained from one more D7-brane without an extra tuning, but 
F-theory picture reveals that this is so only because the monodromy of two-cycles is reduced, 
when they are encapsulated in a larger SO group of the D7--O7 brane system. For this reason, 
F-theory lift of MSSM in intersecting D7-branes satisfies the second criterion of unification.

Let us use $B_3 = \P^1 \times \P^2$, as before, and quantify the number of flux vacua of those 
different implementations of the Standard Model, so that we can compare. Here, we do not pay 
attention to the dimension-4 proton decay problem or any other phenomenological requirements. 
For SU(5) unification, we already have a result, 
\begin{equation}
 h^{3,1} = 3277 \longrightarrow h^{3,1} = 2148, \qquad \Delta h^{3,1} = -1129 \qquad 
  {\rm for~} \SU(5) \quad {\rm on} \quad S = H_{\P^1}.
\end{equation}
If we deform this Calabi--Yau fourfold further so that only $\SU(3) \times \SU(2)$ remains 
unbroken, the two gauge group factors are localised on divisors $S_3$ and $S_2$ both of 
which belong to the same divisor class as $S = H_{\P^2}$.  
\begin{equation}
 h^{3,1} = 3277 \longrightarrow h^{3,1} = 2149, \qquad \Delta h^{3,1} = -1128 \qquad 
  S_3 \sim S_2 \sim H_{\P^1}.
\label{eq:3-2-samediv}
\end{equation}
We can go back to the family of fourfolds for SU(5) unification by suppressing one 
deformation parameter corresponding to $H^0(S; N_{S|B_3}) = H^0(\P^2; {\cal O})$ in this case. 
Therefore, the tuning cost for the unbroken U(1) hypercharge symmetry is obtained by 
$- \Delta h^{3,1} = 1$ in (\ref{eq:estimate-1}) in the case of SU(5) unification.

In case we require SU(3) and SU(2) 7-branes on two divisors, $S_3$ and $S_2$, respectively, 
in different divisor classes in $B_3$, back of the envelope calculation reveals that 
\begin{eqnarray}
 h^{3,1} = 3277 \longrightarrow h^{3,1} =  2155, & \Delta h^{3,1} = -1122 & \qquad  
   S_3 = H_{\P^1}, \quad S_2 = H_{\P^2}, \\
 h^{3,1} = 3277 \longrightarrow h^{3,1} = 2282, & \Delta h^{3,1} = - 995 & \qquad 
   S_3 = H_{\P^2}, \quad S_2 = H_{\P^1}.
\end{eqnarray}
Comparing these $\Delta h^{3,1}$'s with that in (\ref{eq:3-2-samediv}), we see that the 
topological configuration of $\SU(3) \times \SU(2)$ does not make much difference in the 
fraction of flux vacua. If the hypercharge symmetry is obtained as a Mordell--Weil U(1) in 
addition to such $\SU(3) \times \SU(2)$ 7-brane configuration (cf \cite{LW}), $h^{3,1}$ will 
be reduced further by $1000$ or so, as we have experienced in section \ref{ssec:R-parity-violation}.  
The number of flux vacua does not depend very much on topological configuration of 7-branes 
for $\SU(3)_C \times \SU(2)_L$, but it does very much on how we obtain U(1)$_Y$.

How about the tuning cost of $\U(1)_Y$ in the case of intersecting D7-brane system? 
The six D7-branes for $\U(3) \times \U(2) \times \U(1)$ need to be implemented as a part of 
larger D7--O7 system forming an SO group, where 7-brane charges cancel. Thus, minimal tuning 
is not for a rank-4 gauge group, but for a brane system with a rank of gauge group higher 
than 4 in this implementation; the U(1) symmetry of this form is likely to come with a hidden 
tuning cost; it is highly desirable, though, to construct F-theory lift of IIB intersecting 
D7--O7 system, and carry out the same analysis, before drawing a conclusion.   

The original motivation for unification---explaining quantisation of hypercharge---is no longer 
persuasive in string construction of particle physics, because it is explained without relying on 
unification. Unification may still have advantage in F-theory compactification in the geometric 
phase, in that the tuning cost for having an unbroken U(1) hypercharge in addition to 
$\SU(3)_C \times \SU(2)_L$ is small, in terms of flux vacua counting. 

\begin{figure}[tbp]
\begin{center}
\begin{tabular}{ccc}
   \includegraphics[width=.3\linewidth]{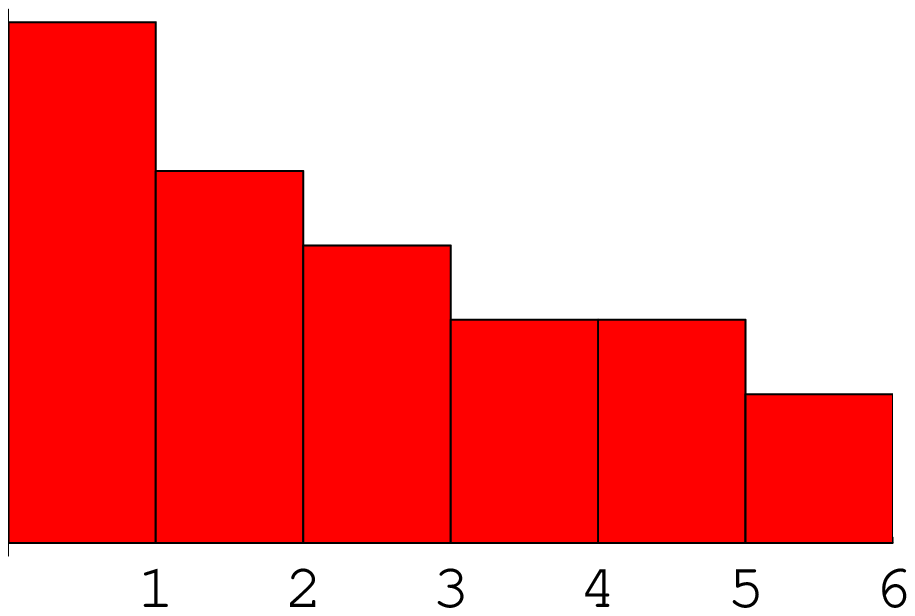}
   & $\qquad$ & 
   \includegraphics[width=.3\linewidth]{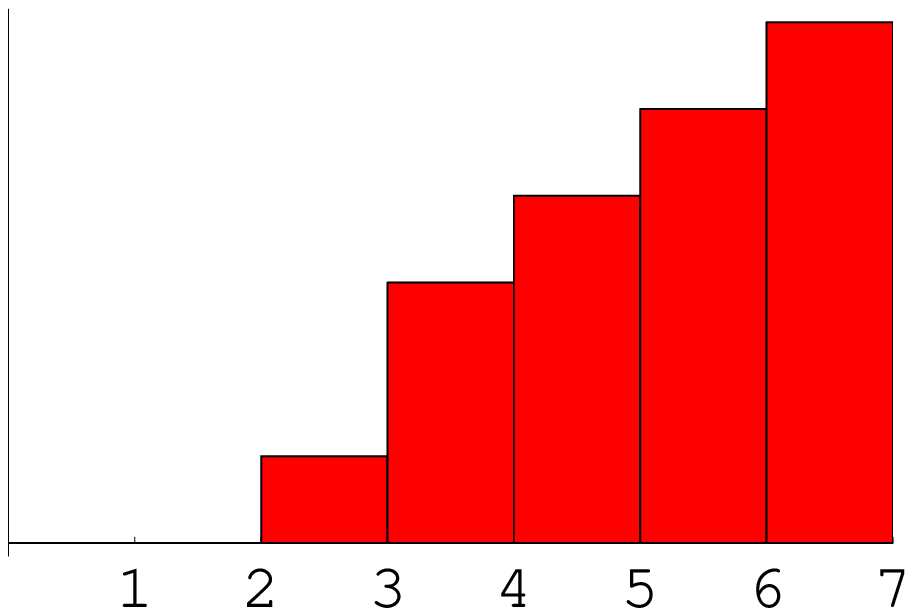}  \\
 (a) $c_1(N_{S|B}) > 0$ ($A_n, D_n$) & & (b) $c_1(N_{S|B}) < 0$ ($A_n, D_n$)
\end{tabular}
\caption{\label{fig:cost-rank} Extra tuning cost $- (\Delta L_*)/(\Delta {\rm rank})$ decreases 
or increases for higher rank, depending on whether $c_1(N_{S|B_3})$ is positive or negative, as 
one goes down the $A_n$ or $D_n$ chain. $c_1(N_{S|B})$ is replaced by $K_S$ in the case of $E_n$ 
series. }
\end{center}
\end{figure}

We should leave a cautionary remark on the $B_3$-dependence of this argument, however. 
Extra tuning cost for one extra rank of 7-brane gauge group has a behaviour shown in 
Figure~\ref{fig:cost-rank}, where the behaviour is qualitatively different for cases with  
``positive'' $c_1(N_{S|B_3})$ and ``negative'' $c_1(N_{S|B_3})$, when one goes down the chain of 
$A_n = \SU(n+1)$ series or of $D_n = \SO(2n)$ series \cite{1408Andreas}.  
When a divisor $c_1(N_{S|B_3})$ on $S$ is negative, in particular, it may happen sometimes that 
$\Delta L_*/\Delta {\rm rank} = 0$ for a choice of 7-brane gauge group with a small rank; the 
rank of 7-brane gauge group can be large to some extent without losing the number of flux vacua. 
This is the phenomenon called non-Higgsable cluster \cite{MV-2,non-Higgsable}. When either 
$\SU(3)_C$ or $\SU(2)_L$ or both are identified with 7-brane gauge groups in a non-Higgsable 
cluster \cite{SM-nonHiggs}, the tuning cost argument above is affected inevitably. 
In a family of fourfolds where the Mordell--Weil group is non-trivial everywhere on its complex 
structure parameter space \cite{always-MW}, we cannot talk of relative statistical cost of 
requiring an extra U(1) symmetry; in such a case, we need to discuss relative tuning cost of 
U(1) through some transitions connecting such a family to another where the fourfolds have 
different topology, or to use the prefactor in (\ref{eq:ADD-formula}) directly to estimate 
the number of flux vacua.

\section{Distribution of Lagrangian Parameters}
\label{sec:distribution}

While the prefactor in the formula (\ref{eq:ADD-formula}) can be used to estimate the number 
of flux vacua with a given set of algebraic and topological properties (i.e., symmetry, matter 
multiplicity etc.), the $(m,m)$-form $\rho_I$ in (\ref{eq:ADD-formula}) can be used to 
``derive'' distribution of Lagrangian parameters in such an ensemble of vacua. This is a source 
of rich information, as is evident already in its applications to Type IIB 
compactifications \cite{taxonomy, CQ}. In this section, we will discuss its F-theory 
applications in the context of particle physics. 

It should be remembered, though, that the expression for $\rho_I$ was derived by assuming that 
the continuous approximation of the $K$-dimensional flux space is good, while the approximation 
is not good in the case of $K \gg L_*$. It may be that the distribution $\rho_I$ remains to 
have reasonable level of predictability, while the problem of bad approximation is mitigated, 
when the complex structure parameter space ${\cal M}_*$ is binned very coarsely, and $\rho_I$ is 
used only by being integrated over such a large bin. Justification is not given even to this 
hope, however. When one is interested in the choice of $(B_3, [S])$ where $K \gg L_*$, 
one should keep this remark in mind. 

\subsection{Symmetry Breaking Scale of an Approximate U(1) Symmetry}
\label{ssec:approx-U1-general}

In section \ref{ssec:approx-U1-general}, we work on an application of the distribution $\rho_I$ 
that extracts its potential power very well. Section \ref{ssec:rank-1} is devoted to a more 
problem-oriented application.  

While it is not theoretically impossible to compute period integrals and evaluate $\rho_I$, 
it is not practical to do so, when there are O(1000) complex structure parameters. Fortunately, 
it is possible to learn essential features of the distribution $\rho_I$ without carrying out 
such computations, as experience in Type IIB applications indicates \cite{taxonomy, ET}. 
First, the parameter space of complex structure ${\cal M}_*$ has a natural set of coordinates;
in the case Calabi--Yau $n$-fold is given by a toric hypersurface, for example, we can use, 
for the coordinates of ${\cal M}_*$, various products of monomial coefficients that are invariant 
under rescaling, \cite{CK}. The distribution $\rho_I$ will show more or less uninteresting 
behaviour 
on these coordinates in ${\cal M}_*$, except at special loci in ${\cal M}_*$. $\rho_I$ exhibits 
singular behaviour in these coordinates, when the period integrals involve logarithm of those 
coordinates; only derivatives of logarithm introduce poles. Logarithm of such coordinates indicates 
that there is a non-trivial monodromy of cycles \cite{ET}. All these arguments above holds true 
for applications to Calabi--Yau fourfolds.\footnote{
Certainly there is small difference between threefolds 
and fourfolds; the number of cycles $b_3 \sim 2h^{2,1}$ for period integrals is not much different 
from $2h^{2,1}$ period integrals forming special coordinates for Calabi--Yau threefolds, there are 
much larger number of four-cycles $b_4 \sim 2h^{3,1} + h^{2,2} \sim 6h^{3,1} + {\rm const}$ than the 
number of independent period integrals for fourfolds. We do not see this difference as a serious 
obstacle in recycling the Type IIB lesson in the main text to the application to F-theory.} 

Consider a family of Calabi--Yau fourfolds $Y_{n=4}$ obtained as a hypersurface of an ambient 
space given by $WP_{[1:2:3]}$-fibration over some $B_3$; 
\begin{equation}
  X^3 + Y^2 + XYZ A_1 + X^2 Z^2 A_2 + Y Z^3 A_3 + X Z^4 A_4 + Z^6 A_6 = 0,
\end{equation}
with $A_k \in \Gamma(B_{n-1=3}; {\cal O}_B(-kK_B))$. Let ${\cal M}_*$ be its parameter space of 
complex structure. Sitting within this family is a family of Calabi--Yau fourfolds with the 
ambient space replaced by ${\rm Bl}_{[1:0:0]}W\P_{[1:2:3]}$-fibration over $B_3$; the last term 
$Z^6 A_6$ is simply dropped (followed by small resolution) to get to the sub-family given 
by (\ref{eq:E7-curve-fibre}), where 
there is a non-trivial section in $Y_4$, and hence a U(1) symmetry in the low-energy effective 
theory. Let ${\cal M}_*^{\U(1)} \subset {\cal M}_*$ be the locus of this sub-family.  
We study the behaviour of $\rho_I$ on ${\cal M}_*$ near the locus of this sub-family. 

In the $A_6 \longrightarrow 0$ limit, $Y_4$ has a curve of codimension-three conifold singularity, 
$X = Y = A_3 = A_4 = 0$ \cite{GW}; this curve in $B_3$ is denoted by $\Sigma$. The conifold transition 
in such a limit was studied extensively in \cite{conifold-CY4}.
The genus of this curve is determined by 
\begin{equation}
  2g(\Sigma)-2 = (-3K_B) \cdot (-4K_B) \cdot (-6K_B) = 72 (c_1(TB_3))^3.
\end{equation}
Incidentally, the parameter space ${\cal M}_*^{\U(1)}$ for this $A_6 \longrightarrow 0$ limit is 
of complex codimension-$(-\Delta h^{3,1})$ in ${\cal M}_*$, where 
\begin{equation}
 (- \Delta h^{3,1}) = h^0(B_3; {\cal O}_B(-6K_B)) - h^0(B_3; {\cal O}_B(-3K_B))
  - h^0(B_3;{\cal O}_B(-2K_B));
\end{equation}
the first term is obviously the degree of freedom in $A_6$. The last two terms are there 
because only the $\epsilon_1$-term in the automorphism of the form 
\begin{equation}
  \delta Y = X Z \epsilon_1 + Z^3 \epsilon_3, \qquad 
  \delta X = Z^2 \epsilon_2, \qquad \epsilon_k \in \Gamma(B_3; {\cal O}_B(-kK_B))
\end{equation}
survives for $Y_4$ in the sub-family over ${\cal M}_*^{\U(1)}$. One can see that 
$(-\Delta h^{3,1})=g$, at least when $B_3$ is a Fano variety. Indeed, because the divisor $(-K_B)$ 
is ample, Kodaira's vanishing theorem implies that 
\begin{equation}
  h^q(B_3; {\cal O}_B(-nK_B)) = 0 \qquad \qquad {\rm for~}q > 0, \quad n \geq 0.
\end{equation}
Combining this theorem and the expression for $(- \Delta h^{3,1})$, we find that 
\begin{equation}
 (- \Delta h^{3,1}) = 36 (c_1(TB_3))^3 + 1 = g.
\end{equation}
This agreement always holds at local level, but $(-\Delta h^{3,1}) = g - \tilde{h}^{2,1} \leq g$ 
at global level \cite{conifold-CY4}; the argument above shows that $\tilde{h}^{2,1} = 0$ at least 
when $B_3$ is a Fano variety. 

$6g-3$ topological four-cycles are identified in the local geometry of $Y_4$ \cite{conifold-CY4}, 
and all of them are lifted to topological cycles in $Y_4$, at least when $B_3$ is a Fano variety. 
Period integrals on these $6g-3$ four-cycles vanish when all the $g$ transverse coordinates of 
${\cal M}_*^{\U(1)} \hookrightarrow {\cal M}_*$ are set to zero; see \cite{conifold-CY4} and the 
appendix \ref{sec:monodromy}. 
We found, in the appendix \ref{ssec:cycle-monodromy}, that there are at least $g$ independent generators 
of unipotent monodromy\footnote{Unipotent monodromy means, here, that a monodromy matrix is a sum 
of a nilpotent matrix and the identity matrix.} acting on these topological four-cycles, and the 
period integrals are of the form,
\begin{equation}
  \Pi_{\widetilde{A}_k} \sim z_k, \qquad \Pi_{\widetilde{C}^k} \sim c^{kl} z_l \ln (z's);
\end{equation}
the $A_6 \longrightarrow 0$ limit corresponds to $z_1=z_2=\cdots z_g = 0$.
It is then quite likely, as in \cite{taxonomy, ET}, that the $(m,m)$-form distribution $\rho_I$ 
on ${\cal M}_*$ has an asymptotic behaviour 
\begin{equation}
 \rho_I \approx \rho_I^{\U(1)} \wedge \rho_I^\perp, \qquad 
 \rho_I^\perp := \prod_{k=1}^g \frac{dz_k \wedge d\bar{z}_k}{|z_k|^2 (\ln(|z|^2))^2}
  \sim \prod_{k=1}^g d [{\rm arg}(z_k)] \wedge \frac{d \ln (1/|z_k|^2)}{[\ln(1/|z_k|^2)]^2}
\end{equation}
near ${\cal M}_*^{\U(1)}$. While derivation of the asymptotic form above is not as rigorous as 
it is desired to be, let us explore what this behaviour implies, if it is true. 

The most important consequence is that the fraction of flux vacua with hierarchically small 
value of $\U(1)$ symmetry breaking parameter $|z_k|$ is not hierarchically small, but is only 
suppressed by some power of the logarithm of the hierarchy, $\ln (1/|z_k|^2)$. That makes it 
much more {\it natural} to think of approximate U(1) symmetry in bottom-up model building. 
Secondly, though, it is likely that the U(1) symmetry is preserved approximately only if all 
the $|z_k|$'s are hierarchically small; that is, what really matters will be a fraction of flux 
vacua satisfying, say, $|z_k|^2 < \delta$ for ${}^\forall k=1,\cdots, g$ for some small $\delta$. 
This then implies that only the fraction 
\begin{equation}
  \int_{{\cal M}_*^{\rm local}; \leq \delta} \rho_I^\perp = 
  \prod_k \left[ \int_{\ln(1/\delta)}^{+\infty} \frac{d \ln(1/|z_k|^2)}{[ \ln (1/|z_k|^2)} \right]  
  = \frac{1}{[\ln(1/\delta)]^g} 
  \label{eq:fraction-approx-U1}
\end{equation}
of flux vacua in ${\cal M}_*$ has such an approximate U(1) symmetry. The value of $g(\Sigma)$ 
is often quite large; when $B_3 = \P^3$, for example, $g = 36 \times 4^3 + 1$. 
Thus, the fraction of flux vacua decreases very quickly, when we require the approximate U(1) 
symmetry to be preserved for very hierarchically small $\delta$. 

Let us take one more step and ask the following question. Although $\rho_I$ 
or (\ref{eq:ADD-formula}) is presented in the form of a continuous distribution, it is originally 
a scatter plot on ${\cal M}_*$ of isolated flux vacua. What is the smallest value $\delta_{\rm min}$ 
of the approximately preserved U(1) symmetry in ${\cal M}_* \; \backslash \; {\cal M}_*^{\U(1)}$? 
This is a prototype of such questions as what the minimum symmetry breaking scale is for 
supersymmetry and flavour symmetry in string landscape.

A wild speculation will be to think as follows. When we set $\delta$ small enough, the fraction 
of flux vacua (\ref{eq:fraction-approx-U1}) becomes so small that it reaches the fraction of 
flux vacua on ${\cal M}_*^{\U(1)}$ among those on ${\cal M}_*$. The integral of $\rho_I^\perp$ over 
the normal coordinates of ${\cal M}_*^{\U(1)} \hookrightarrow {\cal M}_*$ in such a small region 
as $|z_k|^2 < \delta$ may correspond to flux vacua that sit right on top of the 
${\cal M}_*^{\U(1)}$ locus. This thought leads to a relation 
\begin{equation}
  \left[ \frac{1}{\ln(1/\delta_{\rm min})} \right]^g = 
  \exp \left[ \frac{\ln(24)}{4} (\Delta h^{3,1}) \right],
\end{equation}
where (\ref{eq:estimate-1})---valid for cases with $h^{3,1} \gg h^{1,1}$---was used in the right 
hand side. Geometry dependence through $g = -\Delta h^{3,1}$ drops out from this relation then, 
and we find that 
\begin{equation}
 \delta_{\rm min} \sim \exp \left[ - e^{\frac{\ln(24)}{4}} \right].
\end{equation}
This ``prediction'', however, is not as powerful as it looks. We have to keep in mind the limited 
reliability in the value of ``$\ln(24)/4$'', as remarked in footnote \ref{fn:ln-24}. It will not 
be still too bad to conclude that $\delta_{\rm min}$ will not be much smaller than 
\begin{equation}
   \exp[-e^{({\rm a~few})} ] \approx \exp[ - 10 ] \approx 10^{- (3\mbox{-}4)},
\end{equation}
provided all the speculative arguments leading to this conclusion are not wrong. 

\subsection{Statistical Cost of Yukawa Hierarchical Structure Problem}
\label{ssec:rank-1}

In section \ref{ssec:rank-1}, we discuss the fraction of flux vacua that realise solutions to the 
hierarchical structure problem of Yukawa matrices. Each one of codimension-three singularity 
(matter-curve intersection) points in F-theory compactification for SU(5) unification gives rise to 
an approximately rank-1 Yukawa matrix, provided complex structure is generic \cite{Hayashi-Codim, 
HV-flavor, Cecotti, Hayashi-flavor}, but the up-type [resp. down-type and charged lepton] Yukawa 
matrix in the low-energy effective theory below the Kaluza--Klein scale receives contributions from 
all the ``$E_6$''-type points [resp. $D_6$ type] on $S \subset B_3$. The number of ``$E_6$''-type and 
$D_6$-type points are determined by topological intersection numbers, and are generically not equal 
to one \cite{Hayashi-flavor, Cordova}. The approximately rank-1 nature of the Yukawa matrices 
at short distance in F-theory is therefore lost at energy scale below the Kaluza--Klein scale, 
at least in a generic flux vacuum. There have been proposed a few ideas,\footnote{In Heterotic 
string compactification with SU(5) unification, at least some neighbourhoods of orbifold limits 
of the parameter space must be included as a part of the semi-realistic corners of string 
landscape \cite{DESY}. Also, when a Calabi--Yau threefold for compactification has an elliptic 
fibration, one can translate the solutions in F-theory to Heterotic language. The whole picture 
of the landscape of Heterotic string parameter space remains to be far from clear, however. When 
it comes to $G_2$-holonomy compactification of M-theory, the author is unaware of any idea in the 
literature to get around the difficulty in the up-type Yukawa matrix when SU(5) unification 
is assumed \cite{TW-06} (Reference \cite{Bourjaily} arrived at the same observation 
independently).} however, how to exploit the approximate rank-1 nature at short distance. We 
pick up two among them\footnote{In this article, we do not study the statistics of the idea of 
alignment among Yukawa matrices due to a discrete symmetry \cite{Hayashi-flavor}.} for the study 
in this section \ref{ssec:rank-1}.

One of the two ideas is to tune parameters so that only a single ``$E_6$''-type point contributes 
to the up-type Yukawa matrix in the effective theory below the Kaluza--Klein scale (and just one 
$D_6$-type point to the down-type Yukawa matrix); 
this idea was proposed originally in \cite{HV-flavor, Caltech-0904} and the Yukawa matrices in this 
scenario have been studied carefully in \cite{CP, Cecotti, Marchesano-F-flavor}. 
In order to make sure that the low-energy Yukawa matrix receives contribution only from just 
one ``$E_6$''-type point, it is safe to consider that splitting of matter curves is controlled 
by a U(1) symmetry \cite{TTW-RHnu, Hayashi-More, GW}. 

It is true that, for the CKM mixing angles to be small, the single ``$E_6$''-type point and the 
single $D_6$-type point should be at the same point in $S$, or at least be close enough \cite{Randall}; 
this property does not follow from a U(1) symmetry (and matter curve factorisation). If one is happy 
to ignore this aspect in the mixing angle and to focus on the hierarchical structure of the Yukawa 
eigenvalues for now,\footnote{It is understood in phenomenology community, by now, that mixing 
angles will carry more fundamental information than the hierarchical Yukawa eigenvalues (see, 
e.g., \cite{anarchy}). This is because the CKM mixing angles reflect the property only of three 
quark doublets $({\bf 3}, {\bf 2})_{+1/6} \subset {\bf 10}$, and the lepton mixing angles that of 
just the three lepton doublets $({\bf 1}, {\bf 2})_{-1/2} \subset \bar{\bf 5}$, whereas the 
down-type/charged lepton Yukawa eigenvalues reflect the properties of both 
$(\bar{D}, L)= \bar{\bf 5}$ and $(Q,\bar{E}) \subset {\bf 10}$.} then the study in 
section \ref{ssec:R-parity-violation} as well as section \ref{ssec:approx-U1-general} can be used 
to study statistical aspects of this idea of tuning. We will be brief in 
section \ref{sssec:split-mc} for this reason. 

The other idea whose tuning we discuss in section \ref{sssec:large-tau} is a string-theory 
implementation of the idea of \cite{AS}. Sections of a line bundle on a torus $T^2$ (a term 
``magnetised torus'' is sometimes used for this) is given by Theta functions, which become 
approximately Gaussian for large complex structure of $T^2$; the exponentially small tail of the 
Gaussian wavefunctions is used to create hierarchical structure among three copies of 
$(Q,\bar{U},\bar{E}) ={\bf 10}$, which leads to realistic mixing angles and hierarchy in Yukawa 
eigenvalues \cite{anarchy, Gaussian}. See \cite{Madrid-04, Hayashi-flavor} for more detailed 
account of the string implementation of this idea. In this idea, therefore, one assumes that 
the matter curve for $\SU(5)$-{\bf 10} representation has a large complex structure 
parameter.\footnote{Before making this assumption on the complex structure parameter, we make 
another assumption (a discrete choice in topology) that this matter curve has $g=1$. Generalisation 
of this idea to higher genus cases has not been studied very much, apart from partial attempt 
in \cite{Hayashi-flavor}.} We estimate how much fraction of flux vacua we lose by requiring this 
tuning in the complex structure parameter of the matter curve, by exploiting the ``distribution'' 
$\rho_I$.

\subsubsection{Split Matter Curve under a U(1) Symmetry}
\label{sssec:split-mc}

Suppose that the matter curves $\Sigma_{({\bf 10})}$ and $\Sigma_{(\bar{\bf 5})}$ for the ${\bf 10}$
and $\bar{\bf 5}$ representations of Georgi--Glashow SU(5) unification are split into irreducible 
pieces, due to an extra unbroken U(1) symmetry originating from a non-trivial section. 
Let $\Sigma_{({\bf 10})} = \cup_{a} \Sigma_{({\bf 10});a}$ and 
$\Sigma_{(\bar{\bf 5})} = \cup_b \Sigma_{(\bar{\bf 5});b}$ be the irreducible decomposition protected by 
the U(1) symmetry. The idea of \cite{HV-flavor, Caltech-0904} assumes, among other things, that 
there is a pair $\Sigma_{({\bf 10});a0}$ and $\Sigma_{(\bar{\bf 5});b0}$ such that they intersect 
transversely 
(i.e., ``$E_6$''-type) just once in the SU(5) 7-brane locus $S$; the matter 
${\bf 10}=(Q,\bar{U},\bar{E})$ are localised in $\Sigma_{({\bf 10});a0}$ and $H_u$ in 
$\Sigma_{(\bar{\bf 5});b0}$, respectively, so that the single transverse intersection point gives 
rise to the approximately rank-1 up-type Yukawa matrix at low-energy.
It is an interesting question whether there are such Calabi--Yau fourfold geometries. 
The two constructions of fourfolds with a non-trivial Mordell--Weil group which we reviewed in 
the appendix \ref{sec:review-MW} does not have enough freedom to accommodate such configuration 
of matter curves, but this is far from being a no-go. Given the variety of constructions for 
fourfolds with a non-trivial Mordell--Weil group \cite{MW-U1-variety}, it may not be too bad to 
assume that there are constructions satisfying the assumption above. The rest of this 
section \ref{sssec:split-mc} is based on that assumption.  
 
We have already estimated in section \ref{ssec:R-parity-violation} the fraction of flux vacua 
that has an unbroken U(1) symmetry from a non-trivial Mordell--Weil group; factorisation of 
matter curves just follows as a consequence of the U(1) symmetry. Given the fact that the faction 
of such vacua depends on the choice of a construction of fourfolds with a non-trivial Mordell--Weil 
group, as well as on the choice of topology of $(B_3,[S])$, we do not find it meaningful to estimate 
the cost at precision higher than in section \ref{ssec:R-parity-violation}. 
By using the results there, we conclude right away that the cost of extra U(1) symmetry to split 
the matter curves is something like  
\begin{equation}
  e^{-1000} \sim 10^{-{\cal O}(100)} \qquad \qquad 
   {\rm for~} (B_3, [S]) = (\P^1 \times \P^2, H_{\P^1}).
\end{equation}

The tuning cost estimated above should be compared against the naive estimate of the non-triviality 
of flavour structure of the Standard Model, first of all. Suppose that individual Yukawa 
eigenvalues are tuned to be small enough, one by one, by tuning the complex structure parameters 
by hand, and that these tunings for individual eigenvalues are can be carried out independently 
from each other. Then the total tuning cost of the hierarchical eigenvalues of the Standard Model 
by this naive individual tuning is estimated by\footnote{
As we assume SU(5) unification, the hierarchical eigenvalues either in the 
down-type quark sector or charged lepton sector should be taken into account 
in this naive estimate of the tuning, not both. Also, only the ratio of the eigenvalues
is used in this estimate, because the value of $(\tan \beta)$ is not known yet.
On top of the naive estimate in the main text, one should multiply the tuning 
for the small mixing angles in the quark sector, 
$\theta_{us} \cdot \theta_{ub} \cdot \theta_{cb} \sim 10^{-4.5}$, in principle.
We did not include this, however, because the idea of matter-curve splitting 
under the U(1) symmetry does not attempt to reproduce the small CKM mixing angles.} 
\begin{equation}
  \left( \frac{\lambda_c}{\lambda_t} \frac{\lambda_u}{\lambda_t} \right)
  \left( \frac{\lambda_e}{\lambda_\tau} \frac{\lambda_\mu}{\lambda_\tau} \right)
  \approx \left( 10^{-2} \cdot 10^{-4.5} \right) \times 
          \left( 10^{-1} \cdot 10^{-3.5} \right) = 10^{-11}.
\label{eq:brute-force-tuning}
\end{equation}
It is much easier,\footnote{There is no proof, however, that such an accidental tuning for 
individual Yukawa eigenvalues are possible, or impossible, in string theory moduli space.} 
therefore, to obtain the semi-realistic hierarchical structure of Yukawa eigenvalues by just 
an accidental tuning, by chance of $10^{-11}$, than by matter-curve splitting under a 
Mordell--Weil U(1) symmetry, at least for choices of $(B_3, [S])$ with $h^{3,1} \gg h^{1,1}$. 

In fact, we may not have to require that the U(1) symmetry for matter curve splitting is exact. 
Higher precision is required for a U(1) symmetry in the application to the dimension-4 proton 
decay problem, but that is not the case in the application to flavour structure; the level of 
precision required for flavour physics is not more than $m_e/(174 \; \GEV) \sim 10^{-5.6}$. This 
motivates us to pay attention also to flux vacua with an approximate U(1) symmetry, where matter 
curves $\Sigma_{({\bf 10}}$ and $\Sigma_{(\bar{\bf 5})}$ are near the factorisation limit. Qualitative 
aspects of flux vacua distribution with an approximate U(1) symmetry in 
section \ref{ssec:approx-U1-general} will remain the same, even after requiring an extra SU(5) 
symmetry on $S \subset B_3$, because the geometry of U(1) symmetry breaking (i.e., conifold 
transition) along a curve $s^2 a_3 = s^3 a_2 = 0$ in $B_3$ in SU(5) models remains qualitatively 
the same as in the case without SU(5) unification, at least away from the GUT divisor 
$S \subset B_3$. An approximate U(1) symmetry is realised in much larger number of flux vacua 
than an exact U(1) symmetry is, and therefore the tuning problem for the hierarchical structure 
may be alleviated in this way.  

It remains to be seen, however, to what extent the idea of \cite{HV-flavor, Caltech-0904} works 
successfully even in the presence of a small symmetry breaking in the approximate U(1) symmetry. 
The question we asked at the end of section \ref{ssec:approx-U1-general}---the minimum symmetry 
breaking scale $\delta_{\rm min}$---may also become relevant in this context. 

\subsubsection{Gaussian Wavefunction due to Large Complex Structure}
\label{sssec:large-tau}

The second solution to the hierarchical structure problem of low-energy Yukawa matrices also 
requires tuning in one of the complex structure parameters. We use the distribution $\rho_I$ 
in (\ref{eq:ADD-formula}) in order to estimate the fraction of flux vacua for this solution is.

As we have reminded ourselves at the beginning of section \ref{sec:distribution}, the two 
important things in using $\rho_I$ are i) to identify the natural coordinates of the parameter 
space ${\cal M}_*$, and ii) to find out the locus of ${\cal M}_*$ where there is a unipotent 
monodromy on the four-cycles of $\hat{Y}_4$. Although we also need dictionary between the 
coordinates on ${\cal M}_*$ and parameters of the low-energy Lagrangian (Yukawa couplings in 
particular), this part has already been worked out in the literature at the level we need in the 
present context \cite{DW-1, BHV-1, Hayashi-Codim, Hayashi-flavor}.\footnote{Except one caveat: 
see footnote \ref{fn:degeneration-limit}.}  

The dictionary we use is the following. Let us use the base $B_3 = \P^1 \times \P^2$, and the SU(5) 
7-brane locus $S = {\rm pt} \times \P^2 \subset B_3$ for concreteness. With generic choice of complex 
structure of a fourfold $\hat{Y}_4$ for SU(5) unification, the matter curve $\Sigma_{({\bf 10})}$ is 
an irreducible curve\footnote{When the base manifold is 
$B_3 = \P[ {\cal O}_{\P^2} \oplus {\cal O}_{\P^2}(n H_{\P^2})]$, the genus of this matter curve is 
determined by $2g(\Sigma_{({\bf 10})})-2 = (3-n)(-n)$. We chose $n=0$ in this article so that 
$2g-2=0$. } of genus 1, so that we can use the second solution. Let $\tau$ be the complex 
structure parameter of the genus one curve $\Sigma_{({\bf 10})}$. The $j$-invariant of an elliptic 
curve has an expansion 
\begin{equation}
 j \simeq e^{-2 \pi i \tau} + 744 + {\cal O}(e^{ 2\pi i \tau}) 
\end{equation}
which is convenient for large ${\rm Im}(\tau)$. Hierarchical Yukawa eigenvalues as well as small 
mixing angles in the CKM matrix follow, if ${\rm Im}(\tau)$ is parametrically large, or 
equivalently, the value of $|j(\Sigma_{({\bf 10})})|$ is exponentially large. This $j$-invariant 
of the genus one curve should be some modular function over the $m=h^{3,1}=2148$-dimensional space 
${\cal M}_*$ of complex structure of this compactification for SU(5) unification. 

The first task in this section \ref{sssec:large-tau} is to identify the natural coordinates on 
${\cal M}_*$ and to find out how $j(\Sigma_{({\bf 10})})$ depends on these coordinates. 
The Calabi--Yau fourfold $\hat{Y}_4$ in question---for the choice of $(B_3, [S])$---is given as a 
hypersurface of a toric variety: 
\begin{align}
 y^2 + x^3 
+ & (a_{5|0} + s a_{5|1} + s^2 a_{5|2}) xy + (s a_{4|1} + s^2 a_{4|2} + \cdots ) x^2
     \nonumber  \\
+ & (s^2 a_{3|2} + s^3 a_{3|3} + \cdots ) y  + (s^3 a_{2|3} + \cdots ) x
 + (s^5 a_{0|5} + \cdots )= 0,
\end{align}
where $s$ is the inhomogeneous coordinate of $\P^1$, and is regarded as the normal coordinate of 
$S \subset B_3$. We understand here that all the terms corresponding to interior lattice points 
of facets of the dual polytope are set to zero in this defining equation; the automorphism group 
action on the monomial coefficients is now gauge-fixed for the most part, and only the coordinate 
rescaling $(\C^\times)^{4}$ 
acts on the coefficients. As a part of standard story in the toric hypersurface construction of 
Calabi--Yau manifolds, the complex structure parameter space ${\cal M}_*$ is given a natural set of 
coordinates; each one of them  is in the form of 
\begin{equation}
 z_a := \prod_{\tilde{\nu}_\alpha} (a_\alpha)^{\tilde{\ell}^a_\alpha}, 
\label{eq:toric-coordinates}
\end{equation}
where $\alpha$ runs over the monomials in the defining equation, and $a$ labels linear relations 
$\sum_\alpha \tilde{\ell}^a_\alpha \tilde{\nu}_\alpha = 0$ in the dual lattice $M$ of the toric data
(e.g. \cite{CK}). In the case of $\hat{Y}_4$ we consider, there are 2148 such coordinates. 

The matter curve $\Sigma_{({\bf 10})}$ is given by $a_{5|0} = 0$, and $a_{5|0}$ is a cubic 
homogeneous function on $S \cong \P^2$:
\begin{equation}
  a_{5|0} = a^{5|0}_{300} T^3 + a^{5|0}_{210} T^2 U + a^{5|0}_{201} T^2 V + \cdots 
   + a^{5|0}_{003} V^3. 
  \label{eq:gen-cubic}
\end{equation}
None of the ten terms in this cubic form correspond to an interior point of a facet of the dual 
polytope, and hence we should retain all of them. Using the ten coefficients 
$a^{5|0}_{300}, \cdots, a^{5|0}_{003}$, seven independent rescaling invariants (i.e., the coordinates of 
the form (\ref{eq:toric-coordinates})) can be constructed. The $j$-invariant of $\Sigma_{({\bf 10})}$ 
should depend on the seven coordinates out of\footnote{\label{fn:degeneration-limit}
An idea that large ${\rm Im}(\tau)$ of the matter curve $\Sigma_{({\bf 10})}$ results in Gaussian 
profile of wavefunctions along $\Sigma_{({\bf 10})}$ and consequently to hierarchical Yukawa 
eigenvalues is spelled out \cite{Hayashi-flavor} in the language of Katz--Vafa type field theory 
(field theory local model) on $S \times \R^{3,1}$. Very little discussion is found in the 
literature, however, over to what extent we can rely on this field theory picture for 
generic choice of complex structure parameters. Put differently, is it really true that only the 
coefficients $a^{5|0}_{***}$'s are relevant to the hierarchical structure?} the 2148 coordinates 
of ${\cal M}_*$.

Before talking of how the $j$-invariant of a generic cubic curve of $\P^2$ depends on its monomial 
coefficients, let us have a look at the result for easier ones. When an elliptic curve is given in 
the Weierstrass form or Hesse form, the $j$-invariant is given in this way:
\begin{align}
   y^2 = x^3 + f x + g: & \qquad
   j = 4^4 \times 27 \frac{f^3}{4 f^3 + 27 g^2}, \\
    a_1 X^3 + a_2 Y^3 + a_3 Z^3 + a_0 X Y Z = 0: & \qquad  
  j = - \frac{z (z-216)^3}{27 (z + 27)}, \quad 
    z := \left(\frac{a_0^3}{a_1a_2a_3} \right). 
\end{align}
The condition ${\rm Im}(\tau) \gg 1$ corresponds to the vanishing locus of the denominator, 
$4f^3 + 27 g^2 \simeq 0$ or $z + 27 \simeq 0$, or the discriminant locus, to put differently. 
When the defining equation is in the Jacobi form, 
\begin{equation}
  w^2 = c_0 u^4 - c_1 u^3 + c_2 u^2 - c_3 u + c_4, 
\label{eq:Jacobi-form}
\end{equation}
the discriminant locus is given by 
\begin{align}
& 6912 c_0^3 c_4^3  - 3456 c_0^2 c_2^2 c_4^2
 + 432 c_0 c_2^4 c_4 - 5184 c_0^2 c_1 c_3 c_4^2 \nonumber \\
-& 2160 c_0 c_1 c_2^2 c_3 c_4
- 162 c_0 c_1^2 c_3^2 c_4 + 27 c_1^2 c_2^2 c_3^2 - 108 c_1^3 c_3^3 
   \nonumber \\
-& 729 c_1^4 c_4^2 + 3888 c_0 c_1^2 c_2 c_4^2 - 108 c_1^2 c_2^3 c_4 
+ 486 c_1^3 c_2 c_3 c_4 \nonumber \\
-& 729 c_0^2 c_3^4 + 3888 c_0^2 c_2 c_3^2 c_4 - 108 c_0 c_2^3 c_3^2
+ 486 c_0 c_1 c_2 c_3^3 = 0. 
\label{eq:discrim-Jacobi}
\end{align}
For the $j$-invariant of those curves to be exponentially large, which is what we want for 
phenomenology, then the discriminant needs to be exponentially small; that seems to be a general 
lesson from elliptic curves given by those different forms of defining equations. 

The matter curve $\Sigma_{({\bf 10})}$ is given by a generic cubic (\ref{eq:gen-cubic}) in $\P^2$, but 
this is not much different from all the elliptic curves above. Any generic cubic can be cast into 
the Jacobi form (\ref{eq:Jacobi-form}) 
(e.g., \cite{algorithm-cubic}; recent appearance in physics literature includes \cite{Cvetic}).
Using the discriminant locus of the Jacobi form (\ref{eq:discrim-Jacobi}), 
one can then detect the discriminant locus in the coefficients of 
the general cubic form, and hence in the complex structure parameter space 
${\cal M}_*$ of F-theory compactification.
This procedure is easier when such a point as $[T:U:V] = [0:0:1]$ is in 
the curve $a_{5|0}=0$ (i.e., $a^{5|0}_{003}  = 0$); the left-hand side 
of (\ref{eq:discrim-Jacobi})---a homogeneous function of $c_{0,1,2,3,4}$ of 
degree 6---becomes a homogeneous function of $a^{5|0}_{***}$'s ($a^{5|0}_{003}=0$) 
of degree 12. The most general case, where $a^{5|0}_{003}$ does not necessarily 
vanish, can be reduced to the $a^{5|0}_{003}=0$ case above, by redefinition 
of the coordinates, $T \rightarrow T + sol V$, 
$a^{5|0}_{300} sol^3 + a^{5|0}_{201} sol^2 + a^{5|0}_{102} sol + a^{5|0}_{003}=0$.
It appears, then, that the expression (\ref{eq:discrim-Jacobi}) would involve a cubic root of 
a function of the coefficients $a^{5|0}_{***}$'s, but those terms cancel, and the expression of the 
discriminant turns into a form 
\begin{equation}
  \propto {\rm polynomial}_1 \sqrt{ {\rm polynomial}_3 } + {\rm polynomial}_2. 
\end{equation}
The discriminant locus of the general cubic form should be the zero locus 
of an expression proportional to  
$({\rm polynomial}_1)^2 {\rm polynomial}_3 - ({\rm polynomial}_2)^2$.
This polynomial in the ten coefficients $a^{5|0}_{***}$ can be rewritten as a rational function of the seven 
coordinates $z_a$'s of ${\cal M}_*$ modulo an overall factor that is not relevant in the present context.
Once again, this rational function of $z_a$'s needs to be exponentially small, in order for the 
solution to the hierarchical structure problem to work.  

The complex structure parameter space ${\cal M}_*$ has a codimension-1 locus of ${\rm Im}(\tau) = \infty$, 
or equivalently $j(\Sigma_{({\bf 10})}) = \infty$. Unless there is unipotent monodromy around this locus 
(an issue we come back to shortly), the distribution of $\rho_I$ will remain featureless around this 
locus, and the fraction of vacua for the phenomenological solution is estimated by how finely tuned 
the normal coordinate has to be for phenomenology.\footnote{The distribution $\rho_I$ for F-theory 
compactification has been used in this way for phenomenology already in \cite{Hebecker}.}
Hierarchically small Yukawa eigenvalues require that the value of the normal coordinate (the 
rational function in $z_a$'s) be hierarchically small. Because a single tuning of 
$1/j(\Sigma_{({\bf 10})})$ already does the job (including the CKM mixing angles), however, the 
total tuning cost in this solution will not be as severe as $10^{-11}$ (or $10^{-11} \times 10^{-4.5}$) 
estimate for the naive individual tunings in (\ref{eq:brute-force-tuning}).

It is worth noting that the idea of \cite{AS} was to translate the hierarchically small values of 
Yukawa eigenvalues into a moderately large (but not hierarchically large) parameter in the exponent 
(like ${\rm Im}(\tau)$). In the F-theory implementation \cite{Gaussian, Hayashi-flavor} of this 
idea, however, the value of ${\rm Im}(\tau)$ is likely not to be the right measure of required 
fine-tuning, but the value of $1/j(\Sigma_{({\bf 10})}) \sim e^{2 \pi i \tau}$ is, in the statistics of 
F-theory flux vacua, according to the argument above. 

Let us briefly have a look at whether the distribution $\rho_I$ on ${\cal M}_*$ has singularity 
at the $j(\Sigma_{({\bf 10})})= \infty$ locus; if it does, then the right measure of fine-tuning 
will not be $e^{2\pi i \tau}$ but $1/{\rm Im}(\tau)$. Certainly the ${\rm Im}(\tau) = i \infty$ 
point is the locus of unipotent monodromy of one-cycles on $\Sigma_{({\bf 10})}$. There may also be 
some unipotent monodromy among three-cycles in the matter surface for $\SU(5)$-${\bf 10}$ 
representation, because of the monodromy of one-cycles. The matter 
surface---a four-cycle---remains invariant in this limit, however. The author does not have a 
positive or negative evidence for non-trivial monodromy of four-cycles at the 
$j(\Sigma_{({\bf 10})})=\infty$ locus of the complex structure moduli space ${\cal M}_*$; positive 
evidence is necessary in order to avoid the conclusion in the previous paragraph. 

\subsection*{Acknowledgements}

The author thanks Andreas Braun, James Halverson, Bert Schellekens, Yuji Tachikawa, 
Tsutomu Yanagida and Timo Weigand for discussions and communications. He also owes a lot to 
the organisers and 
participants of workshops ``Physics and Geometry of F-theory'' at Max Planck Institute, Munich 
and ``Stringphenomenology 2015'' at IFT, Madrid. This work is supported in part by 
WPI Initiative and a Grant-in-Aid for Scientific Research on Innovative Areas 2303, MEXT, Japan. 

\appendix 

\section{Fourfolds for $\SU(5) \times \U(1)$ Symmetry}
\label{sec:review-MW}

This appendix is a brief summary note on Calabi--Yau fourfold geometry to be used for F-theory 
compactification when one wants to have $\SU(5) \times \U(1)$ symmetry in the effective theory 
below the Kaluza--Klein scale. There may be a few statements in the following that have not 
been written down in the literature, but those results can be derived by using procedure that 
has become almost standard these days. For this reason, only the results are stated, without 
detailed explanation. 

In this article, we only consider elliptic fibration with a section for F-theory compactification; 
let $\hat{Y}_n$ be a non-singular Calabi--Yau $n$-fold, $\pi: \hat{Y}_n \longrightarrow B_{n-1}$ an 
elliptic fibration morphism, and we assume that there is a divisor $\sigma_0$ of $\hat{Y}_n$ which 
is one-to-one with the base $B_{n-1}$ under $\pi$, except in complex codimension-two loci in 
$B_{n-1}$. Low-energy effective theory has a U(1) symmetry, if the elliptic fibration 
$\pi: \hat{Y}_n \longrightarrow B_{n-1}$ has more sections than just a single section 
$\sigma_0$ \cite{MV-2}.  

We restrict our attention to cases where toric surfaces are used to construct the elliptic curve $E$ 
in the fibre. It is best to use a toric surface such as $WP_{[1:2:3]}$ and $F_1 = dP_1$ (Hirzebruch 
surface), where the polytope $\widetilde{\Delta}_F \subset (\Z \oplus \Z) \otimes \R 
=: N_F \otimes \R$ contains a vertex $\nu'_v$ whose two neighbouring lattice points on 
$\widetilde{\Delta}_F$, denoted by $\nu'_a$ and $\nu'_b$, satisfy 
$\nu'_a + \nu'_b = \nu'_v$ \cite{Grassi-toricK3}; the divisor corresponding to $\nu'_v$ then defines 
one point in $E$. In such toric surfaces as ${\rm Bl}_{[1:0:0]}WP_{[1:2:3]}$ and $F_1$ (whose toric data
are shown in Table~\ref{tab:toric-data-fibresurface}), there is one more 
independent divisor which can be chosen to be degree-1 on $E$; 
\begin{table}[tbp]
\begin{center}
\begin{tabular}{cc}
\begin{tabular}{|c|c|}
\hline
 divisor & toric vectors in $N_F$ \\
\hline 
  $D'_X$ & $\nu'_X = (-1,0)$ \\
  $D'_Y$ & $\nu'_Y = (0,-1)$ \\
  $D'_Z$ & $\nu'_Z = (2,3)$ \\
  $D'_W$ & $\nu'_W = (-1,-1)$ \\
\hline
\end{tabular}
& 
\begin{tabular}{|c|c|}
\hline 
 divisor & toric vectors in $N_F$ \\
\hline
  $D'_0$ & $\nu'_0 = (0, 1)$ \\
  $D'_\infty$ & $\nu'_\infty =(0,-1)$ \\
  $D'_1$ & $\nu'_1 =(1,0)$ \\
  $D'_2$ & $\nu'_2 =(-1, 1)$ \\
\hline
\end{tabular}
 \\
 (a) ${\rm Bl}_{[1:0:0]}WP_{[1:2:3]}$ & (b) $F_1$ 
\end{tabular}
\caption{\label{tab:toric-data-fibresurface} 
Toric vectors in $N_F = \Z \oplus \Z$ for ${\rm Bl}_{[1:0:0]}WP_{[1:2:3]}$ (weighted projective space 
$WP^2_{[1:2:3]}$ blown up at one point) and a Hirzebruch surface $F_1 = dP_1$. 
The two neighbouring lattice points of the polytope for $\nu'_Z$ in (a) [resp. $\nu'_0$ in (b)] 
are $(1,2)$ and $(1,1)$ [resp. $\nu'_1$ and $\nu'_2$], which sum up to be $\nu'_Z$ [resp. $\nu'_0$]; 
this means that $D'_Z$ [resp. $D'_0$] can be used as a section. Those two neighbouring points are both 
vertices of the polytope in (b), while they are not in (a); this makes it impossible to introduce 
the twisting by ${\cal O}_{B_{n-1}}(\kappa^{1,2})$ in the case (a) without introducing an unintended 
non-Abelian symmetry $\SU(3) \times \SU(2)$.}
\end{center}
\end{table}
this divisor defines another point in $E$. When such a toric surface is fibred over some base 
$B_{n-1}$ to be an ambient space for $\hat{Y}_n$, those two points in $E$ become sections of the 
elliptic fibration. The rest of this note deals only with the two toric surfaces above. 
See \cite{VB-G-K2} for other choices of toric surfaces to be fibred. 

\setcounter{subsubsection}{0}
\subsubsection{${\rm Bl}_{[1:0:0]}WP_{[1:2:3]}$-fibred Ambient Space}
\label{sssec:BlWP123}

A Calabi--Yau $n$-fold $Y_n$ is constructed as a hypersurface of an ambient space
\begin{equation}
  \P_{\left\{ 
 \begin{array}{lccr} 
      (-1 & 0 & 1 & 1)  \\
      (0 & 1 & 2 & 3 )
 \end{array} \right\} } \left[ {\cal O}_B \oplus K_B \oplus 
     {\cal O}_B \oplus {\cal O}_B \right].
\end{equation}
Here, the rank-4 fibre of the bundle over the base $B_{n-1}$ is made 
projective\footnote{In order not to leave any ambiguity in the notation,  
we remark that the ordinary $WP_{[1:2:3]}$-fibred ambient space for a 
Calabi--Yau with elliptic fibration and a holomorphic section is denoted by 
$\P_{(1 \, 2 \, 3)}[{\cal O}\oplus {\cal O}_B(-2K_B) \oplus {\cal O}_B(-3K_B)] 
= \P_{(1 \, 2 \, 3)}[K_B \oplus {\cal O}_B \oplus {\cal O}_B]$.}
under the $\C^\times \times \C^\times$ action; one can choose two independent 
relations among the toric vectors in the form of $\sum_i \ell_i \nu'_i = 0 
\in N_F = \Z \oplus \Z$, such as $- \nu'_W + \nu'_X + \nu'_Y = 0$ and 
$\nu'_Z + 2 \nu'_X + 3 \nu'_Y = 0$, and define the corresponding $\C^\times$ 
actions as $(\lambda \in \C^\times): X_i \longrightarrow 
X_i \times \lambda^{\ell_i}$ for homogeneous coordinates $X_i$ corresponding to 
the toric divisors $D'_i$. 
This ambient space is a ${\rm Bl}_{[1:0:0]}WP^2_{[1:2:3]}$-fibration over $B_{n-1}$.

A hypersurface $Y_n$ of this ambient space is given by an equation 
\begin{equation}
 X^3 W^2 + Y^2 W + XYZ W A_1 + X^2 Z^2 W A_2 + Y Z^3 A_3 + X Z^4 A_4 = 0,  
\label{eq:E7-curve-fibre}
\end{equation}
where $A_n \in \Gamma(B_{n-1}; {\cal O}_B(-n K_{B}))$ determines the complex 
structure of an elliptic fibred manifold $Y_n$. $X$, $Y$, $Z$ and $W$ are 
the homogeneous coordinates associated with divisors $D_X$, $D_Y$, $D_Z$ and 
$D_W$, respectively, which are the $D'_X$, $D'_Y$ $D'_Z$ and $D'_W$ divisors 
on the fibre, all over the base $B_{n-1}$.
A section $D_Z|_{Y_n}$ is chosen as the zero section $\sigma_0$. Another section 
$\sigma_1 = D_W|_{Y_n} = \{ W=0 \}|_{Y_n}$ does not intersect with the zero section $\sigma_0$.
When the ambient space is blown down to the $WP_{[1:2:3]}$-fibred one, $D_W$ is mapped to 
$(x,y) := (XW/Z^2, YW/Z^3)=(0,0)$.

\begin{figure}[tbp]
\begin{center}
\begin{tabular}{ccc}
  \includegraphics[width=.3\linewidth]{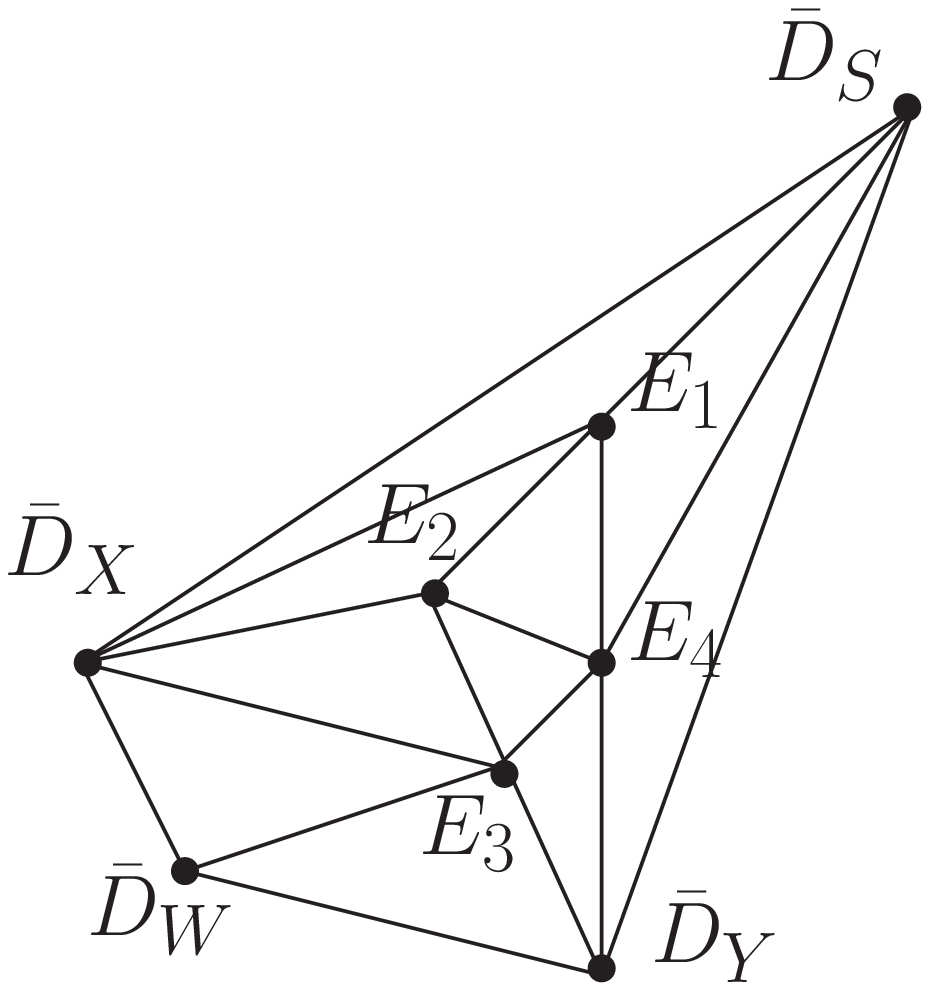}
  & $\qquad \qquad$ &
  \includegraphics[width=.3\linewidth]{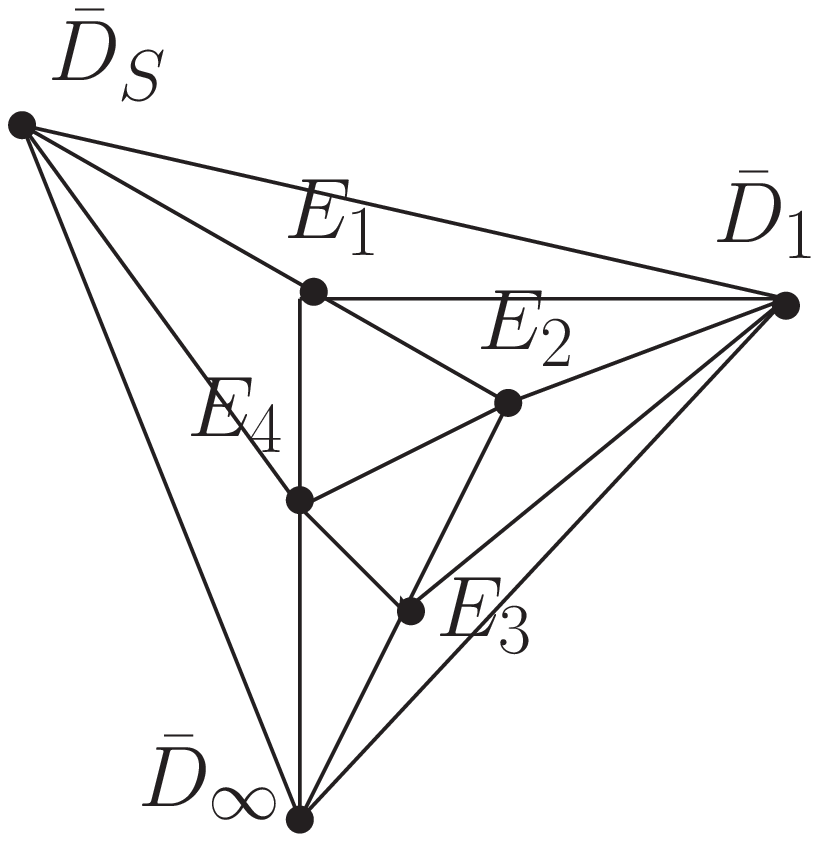}  \\
(a) & & (b)
\end{tabular}
\caption{\label{fig:blow-up}Blow-up procedure shown diagrammatically. 
Subdivision of a triangle using its centre of mass corresponds to a blow-up 
of the ambient space centred at a codimension-three locus, and a subdivision 
of an edge using its centre of mass to a blow-up of the ambient space 
centred at a codimension-two locus. These graphs can be seen as triangulation 
of cones, if the base $B_{n-1}$ is also toric, and the divisor $S$ a toric 
divisor, although we do not assume that $B_{n-1}$ is toric in this summary 
note. The diagram (a) is for the 
${\rm Bl}_{[1:0:0]}WP^2_{[1:2:3]}$-fibred ambient space and (b) for the 
$F_1$-fibred ambient space. Note in (a) that the triangulation of 
$\bar{D}_X$-$E_3$-$\bar{D}_Y$-$\bar{D}_W$ resolves the conifold singularity 
associated with the $\U(1)$-charge $\pm 5$ matter field; the graph (b) is 
the same as the blow-up procedure in \cite{GH}. }
\end{center}
\end{figure}

The $n$-fold $Y_n$ develops a complex codimension-two locus of $A_4$ singularity (when the fibre 
of the ambient space is blown down to $WP_{[1:2:3]}$), when we require 
\begin{equation}
 A_n = s^{n-1} a_{6-n}, \qquad 
    a_{n-1} \in \Gamma(B_{n-1}; {\cal O}_B(-n K_B-(n-1)S)) 
\end{equation}
for $n=1,\cdots, 4$. $S$ is a divisor of $B_{n-1}$, and $s$ is a section of ${\cal O}_B(S)$ such 
that $S = \{ s=0 \}$. A non-singular $\hat{Y}_n$ is constructed by a standard process of $A_4$ 
singularity resolution, followed by small resolutions associated with loci of charged matter 
fields; Figure~\ref{fig:blow-up}~(a) describes this process diagrammatically.\footnote{
$\nu'_{E1} = \nu'_S + \nu'_X + \nu'_Y$, $\nu'_{E2} = \nu'_{E1} + \nu'_X + \nu'_Y$, 
$\nu'_{E4} = \nu'_{E1} + \nu'_Y$, $\nu'_{E3} = \nu'_{E2} + \nu'_Y$. Then add 
$\nu'_W$. A 1-simplex (2-dim cone) $<\nu'_W \nu'_{E3}>$ bisecting the cone 
$<\nu'_{E3}\nu'_X\nu'_Y>$ provides a small resolution of the conifold singularity over the 
$a_2 = a_3=0$ locus in $B_{n-1}$.} Let this blow-up morphism 
be $\nu: \hat{Y}_n \longrightarrow Y_n$; we also use the same $\nu$ for 
the morphism between the corresponding ambient spaces. 

The zero section of $(\pi \cdot \nu): \hat{Y}_n \longrightarrow B_{n-1}$ is given by 
$\sigma_0 := \nu^*(D_Z)|_{\hat{Y}_n}$; we will drop ``$\nu^*$'' or ``$|_{\hat{Y}_n}$'' in the following 
for simpler notations, however, unless subtleties are involved. Another section $\sigma_1 \sim D_W$ 
for $\pi: Y_n \longrightarrow B_{n-1}$ defines a section in $\hat{Y}_n$ except subtleties in 
the fibre of $S \subset B_{n-1}$. When we set 
\begin{equation}
  \sigma''_1 \sim \bar{D}_W - D_Z + K_B + (2E_1 + 4 E_2 + 6 E_3 + 3E_4)/5, 
\label{eq:U1-generator-BlWP123-case}
\end{equation}
where $\bar{D}_W$ is the proper transform of $D_W$ under 
$\nu: \hat{Y}_n \longrightarrow Y_n$, and $E_{1,2,3,4}$ the four exceptional divisors of 
$\nu: \hat{Y}_n \longrightarrow Y_n$, all of $\sigma''_1 \cdot \sigma_0$ and 
$\sigma''_1 \cdot E_{1,2,3,4}$ are mapped to the trivial divisor class in $B_{n-1}$ under 
$(\pi \cdot \nu)_*$. References for the statements up to this point include \cite{Pantev, GW, 
Morrison-Park}. 

There are three distinct groups of $\SU(5)$-charged matter fields in this 
case \cite{Hayashi-More}, as summarised in Table~\ref{tab:ch-matter-summary-WP123}.
\begin{table}[tbp]
\begin{center}
\begin{tabular}{|c|c|ccc|}
\hline 
 bdle & repr. & curve def. eq. & curve div. class & vanishing cycle \\
\hline 
${\bf 3}$ & ${\bf 10}_{-1}$ &  $a_5|_S = 0$ & $(-K_B)|_S$ &
        $(-E_2 \cdot E_4)|_{\hat{Y}_n}$ \\
$\wedge^2 {\bf 3}$ & $\bar{\bf 5}_{-2}$ &  $(a_4 a_3 - a_2 a_5)|_S = 0$ & 
   $(-3S -5K_B)|_S$ & $\left(-(\bar{D}_Y-K_B-E_4)\cdot E_3\right)|_{\hat{Y}_n}$ \\
$\wedge^3 \bar{\bf 3}$ & $\bar{\bf 5}_{+3}$ &  $a_3|_S = 0$ & 
  $(-2S-3K_B)|_S$ & $(\bar{D}_X \cdot E_3)|_{\hat{Y}_n}$ \\
\hline
\end{tabular}
\caption{\label{tab:ch-matter-summary-WP123}
Summary of geometry associated with the SU(5)-charged matter fields in 
the case of ${\rm Bl}_{[1:0:0]}WP^2_{[1:2:3]}$-fibred ambient space.
The matter locus---codimension-1 in $S$---is given by the defining equation 
in the third column; this matter locus belongs to the divisor class on 
$S$ shown in the fourth column. The last column shows the corresponding class 
of vanishing cycle (complex codimension-two in $\hat{Y}_n$). The first column 
shows the representation of the $\U(3)$ structure group of the Higgs bundle 
on $S$. }
\end{center}
\end{table}
The U(1)-charge of these $\SU(5)$-charged matter fields can be determined by using the topological 
class of $\sigma''_1$ in (\ref{eq:U1-generator-BlWP123-case}); the results---shown in 
Table~\ref{tab:ch-matter-summary-WP123}---indicates that the U(1) symmetry generated by $\sigma''_1$ 
shows up as the U(1) part of the $\U(3) \subset E_7$ structure group of the Higgs bundle in the 
field theory local model (Katz--Vafa type field theory) on $S \times \R^{3,1}$ 
(cf \cite{Hayashi-More, GW}). The 6D anomaly cancellation condition indicates that an SU(5)-neutral 
hypermultiplet with U(1)-charge $\pm 5$ is localised in the fibre of a codimension-two 
$a_2 = a_3 = 0$ locus in $B_{n-1}$, and that they are all the matter fields charged under the 
$\SU(5) \times \U(1)$ symmetry (see \cite{GW, Taylor-anomaly}). 

\begin{figure}[tbp]
\begin{center}
\begin{tabular}{ccc}
  \includegraphics[width=.22\linewidth]{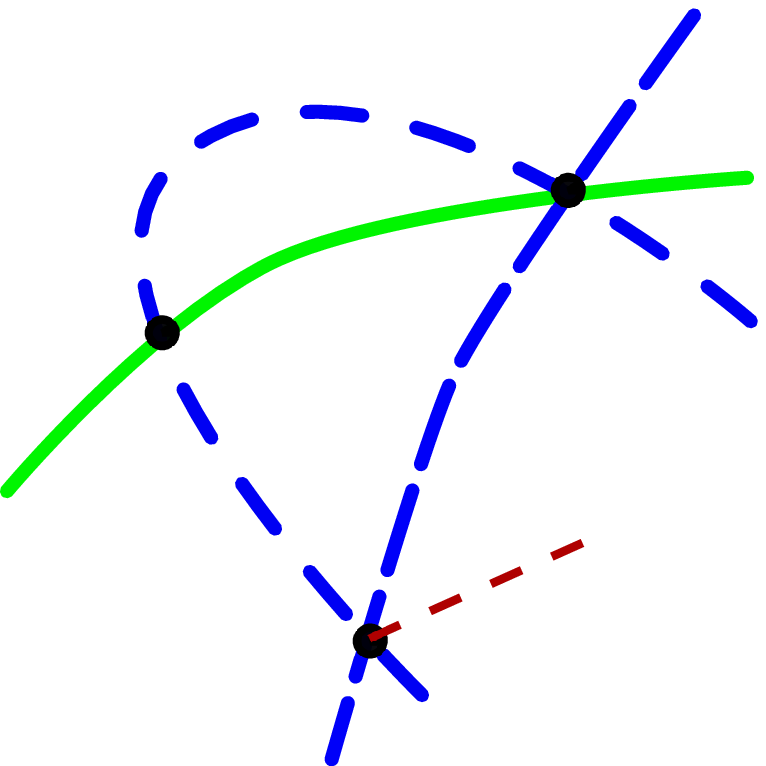}
  & $\qquad$ & 
  \includegraphics[width=.3\linewidth]{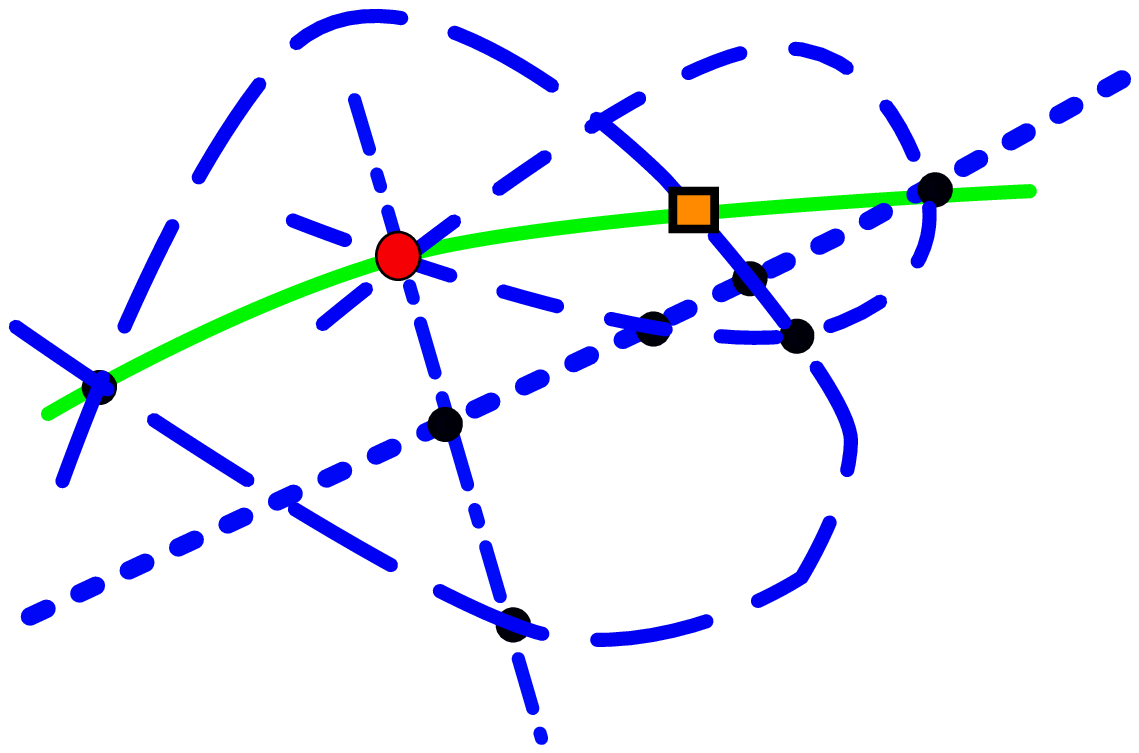}  \\
(a) & & (b)
\end{tabular}
\caption{\label{fig:matter-curve-geom}(colour online) Configuration of matter curves and 
interaction points on the SU(5) 7-brane $S$ shown schematically. The picture (a) is for the case 
of ${\rm Bl}_{[1:0:0]}WP^2_{[1:2:3]}$-fibred ambient space, and (b) for the case of $F_1$-fibred ambient 
space with the no.2 choice of the order of vanishing. Solid curve (green) is the matter curve for 
$\SU(5)$-{\bf 10} representation in both (a) and (b). In the picture (a), the long dashed and 
dashed curves (both blue) are the matter curves for $\bar{\bf 5}_3$ and $\bar{\bf 5}_{-2}$,
respectively. The dotted curve (red) is where SU(5)-neutral U(1)-charged fields are 
localised in $B_3$, ``projected'' on to $S$. In the picture (b), the long dashed, dashed, dotted 
and dash-dotted curves (all in blue) are the matter curves for $\bar{\bf 5}_0$, $\bar{\bf 5}_{-1}$, 
$\bar{\bf 5}_{1}$ and $\bar{\bf 5}_{2}$ representations, respectively. The ``$E_6$''-type 
point for up-type Yukawa is indicated by a square (orange), while the 
point with $F_1$-fibre by a large circle (red). 
}
\end{center}
\end{figure}

This construction can be used for spontaneous R-parity violation. The hierarchical structure 
problem of Yukawa eigenvalues, however, cannot be solved by using this construction 
(without further symmetry or tuning of parameters), because all the ``$E_6$''-type points on $S$ 
contribute to the up-type Yukawa matrix in the low-energy effective theory. See 
Figure~\ref{fig:matter-curve-geom}~(a) for the configuration of matter curves. 

\subsubsection{$F_1$-fibred Ambient Space}
\label{sssec:F1}

$F_1 = dP_1$ can be used as fibre of the ambient space, instead of ${\rm Bl}_{[1:0:0]}WP^2_{[1:2:3]}$, 
in constructing a Calabi--Yau $n$-fold with a non-trivial Mordell--Weil group. We then use an ambient 
space 
\begin{equation}
 \P_{ \left\{ \begin{array}{lccr}
      (1 & 1 & 0 & 0) \\
      (0 & 1 & 1 & 1) \end{array} \right\} } 
 \left[ K_B \oplus {\cal O}_B \oplus 
       {\cal O}_B(\kappa^1) \oplus {\cal O}_B(\kappa^2) \right],
\end{equation}
where the fibre can be twisted by introducing two divisors $\kappa^1$ and $\kappa^2$ of the base 
$B_{n-1}$ \cite{CF}. The fibre is $F_1$; the four line bundles above correspond to the toric vectors 
$\nu'_0$, $\nu'_\infty$, $\nu'_1$ and $\nu'_2$ in Table~\ref{tab:toric-data-fibresurface}~(b), 
respectively. The zero locus of the line bundles are the divisors denoted by $D_{0,\infty,1,2}$, 
and the corresponding homogeneous coordinates are denoted by $X_{0,\infty, 1,2}$.
There are linear equivalence relations 
\begin{equation}
  D_1 - \kappa^1 \sim D_2 - \kappa^2, \qquad 
  D_\infty \sim D_0 - K_B + D_2-\kappa^2.
\end{equation}

An elliptic fibred Calabi--Yau $n$-fold $Y_n$ is given as a hypersurface 
of this ambient space by\footnote{This equation can also be written down 
by using Affine charts for the fibre.
In the chart corresponding to a cone $\vev{\nu'_0, \nu'_1}$ [resp. 
$\vev{\nu'_0,\nu'_2}$], Affine coordinates are 
$(u,\omega) = (X_1/X_2,X_2X_0/X_\infty)$ [resp. 
$(v,\omega') = (X_2/X_1,X_1X_0/X_\infty)$]. In the chart for the cone 
$\vev{\nu'_\infty, \nu'_1}$ [resp. $\vev{\nu'_\infty, \nu'_2}$], the Affine 
coordinates are $(u,w) = (X_1/X_2, X_\infty/(X_0X_2))$ [resp.
$(v,w') = (X_2/X_1, X_\infty/(X_0X_1))$].} 
\begin{align}
&
 X_\infty^2 (A_{0,1} X_1 + A_{1,0} X_2) +
 X_\infty X_0 (B_{-1,1}X_1^2 + B_{0,0} X_1 X_2 + B_{1,-1}X_2^2)  \nonumber \\
&
 + X_0^2 (C_{-2,1}X_1^3 + C_{-1,0}X_1^2X_2 + C_{0,-1}X_1X_2^2 + C_{1,-2}X_2^3) = 0.
\label{eq:defeq-F1-fibre}
\end{align}
Complex structure of $Y_n$ is encoded in the choice of 
\begin{align}
 A_{n_1, n_2} & \in \Gamma(B_{n-1}; {\cal O}(n_1 \kappa^1 + n_2 \kappa^2)), 
     \nonumber \\ 
 B_{n_1, n_2} & \in \Gamma(B_{n-1}; {\cal O}(n_1 \kappa^1 + n_2 \kappa^2 - K_B)),
  \label{eq:def-ABC-F1fibre-gen} \\ 
 C_{n_1, n_2} & \in \Gamma(B_{n-1}; {\cal O}(n_1 \kappa^1 + n_2 \kappa^2 - 2K_B)).
    \nonumber  
\end{align}

We take $D_0$ ($X_0 = 0$ locus) as the zero section\footnote{It is a rational section, but not 
a holomorphic one, when $\kappa^1 \cdot \kappa^2$ is non-empty.}
$\sigma_0$ of the elliptic fibration $\pi_Y: Y_n \longrightarrow B_{n-1}$. 
There is also a section corresponding to the degree-1 divisor $(D'_1-D'_0)|_E$ 
of the fibre, which is denote by $\sigma_1$. It is geometrically given by 
\begin{align}
 [ X_\infty (B_{-1,1}X_1^2 + \cdots + B_{1,-1} X_2^2) +
   X_0 (C_{-2,1}X_1^3 + \cdots + C_{1,-2}X_2^3) = 0 ] - 2 [ X_\infty = 0 ], 
\end{align}
and belongs to the divisor class $(D_1 - D_0 + \kappa^2)$. Since 
\begin{align}
&
 \pi_{Y*}(\sigma_1 \cdot \sigma_0) = [B_{sym} = 0], \\
&
 B_{sym} := B_{1,-1} A_{0,1}^2 - B_{0,0} A_{0,1} A_{1,0} + B_{-1,1} A_{1,0}^2 \in 
   \Gamma \left( B_{n-1}; {\cal O}(\kappa^1 + \kappa^2 - K_B) \right), 
\end{align}
we take 
\begin{equation}
 \sigma''_1 := 
 \sigma_1 - [B_{sym}=0] - \sigma_0 + K_B \sim (D_1 - \kappa^1 - 2 D_0 + 2 K_B)
\end{equation}
as the generator of a U(1) symmetry in the low-energy effective theory.

The charge-$\pm 2$ matter fields under this U(1) symmetry are localised 
in the codimension-two locus of $B_{n-1}$ given by\footnote{
Consider the case $X_n$ is a threefold. In the $I_2$ fibre of a 
such a codimension-2 point in the base $B_2$, $\sigma_0$ is a point in 
one of the two $\P^1$'s, and $\sigma_1$ wraps that $\P^1$. In the $I_2$ 
fibre over a $A_{1,0} = A_{0,1}=0$ point, however, $\sigma_0$ wraps 
one of the two $\P^1$'s (being a rational section when 
$\kappa^1 \cdot \kappa^2$ is non-empty), while $\sigma_1$ wraps 
the other $\P^1$. } 
\begin{align}
 B_{sym} &= C_{sym} = 0, \\
 C_{sym} &:= C_{1,-2} A_{0,1}^3 - C_{0,-1} A_{0,1}^2 A_{1,0} 
          + C_{-1,0} A_{0,1} A_{1,0}^2 - C_{-2,1} A_{1,0}^3 
  \in \Gamma\left( B_{n-1}; {\cal O}(\kappa^1 + \kappa^2 - 2 K_{B}) \right).
    \nonumber  
\end{align}
Matter fields with charge $\pm 1$ are localised in a class 
$4(\kappa^1 + \kappa^2 - 2K_B) \cdot (-\kappa^1 - \kappa^2 - 2 K_B)$.
All that has been stated so far is the same as (or obvious generalisation 
of) \cite{Morrison-Park, VB-G-K}.

Let us consider a case where an $n$-fold $Y_n$ develops $A_4$ singularity at the $X_\infty = X_1 = 0$ 
point in the $F_1$ fibre over a divisor $S \subset B_{n-1}$, so that there is a stack of 7-branes 
for an SU(5) gauge theory along $S \subset B_{n-1}$. The sections $A_{n_1, n_2}$, $B_{n_1, n_2}$ and 
$C_{n_1, n_2}$'s defining the complex structure of the $n$-fold $Y_n$ need to have certain order 
of vanishing along the divisor $S \subset B_{n-1}$ then. There are a couple of different choices, 
as shown in Table~\ref{tab:Tate-A4}, at least in a study of local geometry. 
\begin{table}[tbp]
\begin{center}
\begin{tabular}{|c|cc|ccc|cccc|}
\hline
 choice
 & $A_{0,1}$ & $A_{1,0}$ & $B_{-1,1}$ & $B_{0,0}$ & $B_{1,-1}$ & 
   $C_{-2,1}$ & $C_{-1,0}$ & $C_{0,-1}$ & $C_{1,-2}$ \\
\hline
no.1 & 0 & 0 & 0 & 0 & 1 & 2 & 3 & 4 & 5 \\
no.2 & 0 & 0 & 0 & 0 & 2 & 0 & 1 & 3 & 5 \\
no.3 & 0 & 1 & 0 & 0 & 3 & 0 & 0 & 2 & 5 \\
no.4 & 0 & 3 & 0 & 0 & 4 & 0 & 0 & 1 & 5 \\
\hline 
\end{tabular}
\caption{\label{tab:Tate-A4}
The order of vanishing required for $A_4$ singularity. }
\end{center}
\end{table}
The no.3 choice of the order of vanishing, however, may have a problem, when a global geometry is 
studied; at least in a few examples using compact toric ambient spaces, we found that the 
singular fibre over $S$ in a resolved $n$-fold $\hat{Y}_n$ becomes $I_6$ type of Kodaira 
classification unintentionally. The rest of this summary note focuses on the no.2 choice of the 
order of vanishing. 
It is not clear whether the choice of toric vectors in section 3 of \cite{VB-G-K} 
corresponds to any one of the order of vanishing in Table~\ref{tab:Tate-A4}.

Under the no.2 choice of the order of vanishing, singular $Y_n$ can be made non-singular (denoted 
by $\hat{Y}_n$) by successive blow-ups of the ambient space; the same blow-up procedure as 
in \cite{EY}, shown in Figure~\ref{fig:matter-curve-geom}~(b), does the job in this case. 
The proper transforms of the divisor $D_1$, $D_\infty$ and $D_S = \pi_Y^*(S)$ are denoted by 
$\bar{D}_1$, $\bar{D}_\infty$ and $\bar{D}_S$, respectively. 
\begin{equation}
D_S = \bar{D}_S + E_1 + E_2 + E_3 + E_4, \quad 
D_1 = \bar{D}_1 + E_1 + 2E_2 + 2E_3 + E_4, \quad 
D_\infty = \bar{D}_\infty  + E_1 + 2E_2 + 3E_3 + 2E_4.  
 \nonumber 
\end{equation}
When we choose 
\begin{equation}
 \sigma''_1 \sim (D_2 - \kappa^2 - 2 D_0 + 2K_B)
\end{equation}
as a U(1) generator, the conditions 
$(\pi_Y \cdot \nu)_*(\sigma_0 \cdot \sigma''_1)=(\pi_Y \cdot \nu)_*(E_{1,2,3,4} \cdot \sigma''_1)=0 
\in {\rm Pic}(B_{n-1})$ are satisfied. 

SU(5)-charged matter fields are localised in five distinct codimension-1 loci 
in $S$, as summarised in Table~\ref{tab:ch-matter-summary-F1-no2}.
\begin{table}[tbp]
\begin{center}
\begin{tabular}{|c|ccc|}
\hline
 & matter curve def. eq. ($|_S = 0$) & curve divisor class ($|_S$)
 & vanishing cycle ($|_{\hat{Y}_n}$) \\
\hline
 ${\bf 10}_0$ & $B_{0,0}$ & $-K_B$ & $- E_2 \cdot E_4$ \\
 $\bar{\bf 5}_0$ & 
    $(c_{1,-2|5}B_{0,0}^2 - c_{0,-1|3}B_{0,0}b_{1,-1|2}+c_{-1,0|1}b_{1,-1|2}^2)$ &
    $\kappa^1 - 2 \kappa^2 - 4 K_B - 5S$ & 
    $E_3 \cdot (\bar{D}_1-E_2 - K_B)$  \\ 
 $\bar{\bf 5}_{-1}$ &   $A_{0,1}B_{0,0}^2 - A_{1,0}B_{-1,1}B_{0,0} + C_{-2,1}A_{1,0}^2$ &
     $\kappa^2-2K_B$ & $-\bar{D}_S \cdot (D_0 - E_1-K_B)$ \\
 $\bar{\bf 5}_1$ & $C_{-2,1}$ & $-2\kappa^1 + \kappa^2-2K_B$
     & $\bar{D}_S \cdot \bar{D}_\infty$ \\
 $\bar{\bf 5}_2$ &  $A_{1,0}$ &  $\kappa^1$ & 
     $\bar{D}_S \cdot (\kappa^1-\bar{D}_1)$  \\
\hline
\end{tabular}
\caption{\label{tab:ch-matter-summary-F1-no2}
Summary of geometry associated with SU(5)-charged matter fields in the case of 
$F_1$-fibred ambient space, and the no.2 choice of the order of vanishing. 
See caption of Table \ref{tab:ch-matter-summary-WP123}.}
\end{center}
\end{table}
There, we used the following notations, as in \cite{6authors, Caltech-0904}:
\begin{equation}
 B_{1,-1} =: s b_{1,-1|1}, \quad C_{-1,0} =: s c_{-1,0|1}, \quad 
 C_{0,-1} =: s^3 c_{0,-1|3}, \quad C_{1,-2} =: s^5 c_{1,-2|5}. 
  \label{eq:def-ABC-F1fibre-A4}
\end{equation}
The divisor classes of $\bar{\bf 5}$-representation matter fields 
sum up to be $(-8K_B-5S)|_S$, which is vital to the 6D box anomaly cancellation.
There are also SU(5)-neutral, but U(1)-charged matter fields. 
Their location---codimension-two in $B_{n-1}$---is inferred by using 
the 6D anomaly cancellation conditions; we are led to the following 
solution:
\begin{eqnarray}
 {\rm charge}~\pm 2 & & (\kappa^1 + \kappa^2 - K_B) \cdot 
     (\kappa^1+\kappa^2-2K_B) - 5 S \cdot \kappa^1 \subset B_{n-1}, \\
 {\rm charge}~\pm 1 & & 16K_B^2 - 4(\kappa^1 + \kappa^2)^2 
     - 10 S \cdot (-\kappa^1 + \kappa^2 - 2K_B) \subset B_{n-1}.
\end{eqnarray}
A part of the $B_{sym} = C_{sym}=0$ locus for the charge-$\pm 2$ 
fields---$5 S \cdot \kappa^1$---has been subtracted, which is reasonable 
because the $B_{sym}=C_{sym}=0$ conditions are satisfied automatically at 
$A_{1,0} = s = 0$.

When F-theory is compactified to 3+1-dimensions in this way, by using 
a Calabi--Yau fourfold $\hat{Y}_{n=4}$, geometric configuration of the matter 
curves on $S$ is schematically like Figure~\ref{fig:matter-curve-geom}~(b). 
Most of the intersection points of the matter curves in $S$ are one of the ``$E_6$''-type, 
$D_6$ type and $A_6$-type, but none of those local descriptions apply to the intersection points 
where matter curves for ${\bf 10}_0$, $\bar{\bf 5}_2$ and $\bar{\bf 5}_{-1}$-representations meet. 
The fibre of $(\pi_Y \cdot \nu): \hat{Y}_{4} \longrightarrow B_{3}$ is a surface $F_1$ at 
such points in $B_{3}$. Tensionless strings may show up in the effective theory on 3+1-dimensions 
in this case \cite{tensionless}. For phenomenological purposes, it is thus safe to restrict our 
attention to cases where the divisor class $\kappa^1|_S$ is trivial (so that $A_{1,0}|_S$ remains 
non-zero on $S$). 
 
This $\kappa^1|_S=0$ condition implies, first of all, that the $\bar{\bf 5}_2$--${\bf 5}_{-2}$ 
matter fields do not appear in the low-energy spectrum. When this set-up with a U(1) symmetry 
is used for spontaneous R-parity violation scenario, matter identification should be the following.
First, the up-type Higgs needs to be identified with the doublet part of ${\bf 5}_0$ so that the 
up-type Yukawa couplings are generated. Secondly, for the charged lepton Yukawa couplings to be 
generated, $L$ and $H_d$ need to originate from $\bar{\bf 5}_1$ and $\bar{\bf 5}_{-1}$ or vice versa.
$\bar{D}$'s of the supersymmetric Standard Models need to be on the same matter curve as $L$'s 
in order for the down-type Yukawa couplings to be generated. 

The $\kappa^1|_S=0$ condition also implies that the splitting of the matter curve of $\bar{\bf 5}$ 
representation in this set-up cannot be used for the hierarchical structure problem of the up-type 
Yukawa matrix. In the absence of the matter curve of $\bar{\bf 5}_2$ matter field and of the 
interaction points indicated by a large circle (red) in Figure~\ref{fig:matter-curve-geom}~(b),
all the ``$E_6$'' type points arise in the form of ${\bf 10}_0$--${\bf 10}_0$--$\bar{\bf 5}_0$ 
interaction points at $c_{-1,0|1}=B_{0,0}=0$. Therefore, the result of \cite{Hayashi-flavor, Cordova} 
that the number of ``$E_6$''-type points is even still holds true.

\section{SU(6) 7-brane for Up-type Yukawa Coupling}
\label{sec:SU(6)}

Reference \cite{TW-06} introduced a class of F-theory compactification 
with a stack of SU(6) 7-branes at the divisor $S$ in the base $B_{n-1}$, 
which accommodates SU(5) unification and generates its up-type Yukawa 
couplings. Some details of the construction of this class of compactification 
were missing in \cite{TW-06}, however. Thanks to the development in the 
study of F-theory since then, we can fill the missing details now. 

Let us first note that the class of F-theory compactification with 
an SU(6) 7-brane locus above is somewhat different from general F-theory 
compactification characterised by the Tate condition for the $I_6$-type 
singular fibre. To see this, remember that the Tate condition for the 
$I_6$-type singular fibre in a non-singular elliptic fibration 
$\pi: \hat{Y}_n \longrightarrow B_{n-1}$ corresponds to the following set 
of the order of vanishing of the coefficients in the generalised 
Weierstrass form \cite{6authors}:
\begin{eqnarray}
 0 &=& y^2 + x^3 + A_1 xy + A_2 x^2 + A_3 y + A_4 x + A_6, \\
 && A_1 \sim s^0, \quad A_2 \sim s^1, \quad A_3 \sim s^3, \quad 
    A_4 \sim s^3, \quad A_6 \sim s^6;
\end{eqnarray}
here, $s \in \Gamma(B; {\cal O}_B(S))$, and the $\{s=0\}$ locus corresponds 
to the divisor $S$. It is thus convenient to write the Weierstrass equation 
in the following form:
\begin{equation}
  y^2 + x^3 + a_{5} xy + a_{4} s x^2  + a_{3} s^3 y + a_{2} s^3 x + a_{0} s^6 = 0.
\end{equation}
where $a_{0,2,3,4,5}$ are holomorphic sections of appropriate line bundles on 
$B_{n-1}$. 

When we consider F-theory compactification of this type to 3+1-dimensions, 
using a Calabi--Yau fourfold, straightforward analysis reveals that 
the matter curves in $S$ are given by 
\begin{equation}
  \Sigma_{(\wedge^2 {\bf 6})}: \quad a_5|_S = 0, \qquad 
  \Sigma_{({\bf 6})}: \quad (a_2^2 - a_2 a_5a_3 + a_0 a_5^2)|_S = 0; 
\end{equation}
Katz--Vafa type field theory for these matter fields are SO(12) ($D_6$) and SU(7) ($A_6$) 
gauge theories, respectively. These two matter curves intersect at points $a_5|_S = a_2|_S = 0$; 
physics around these points (including Yukawa couplings) is captured by a field theory 
with SO(14) ($D_7$) gauge group. We cannot expect an up-type Yukawa 
coupling of the form $\Delta W  \sim {\bf 10}^{\cdot \cdot} {\bf 10}^{\cdot \cdot} 
{\bf 5}^{\cdot} \epsilon_{\cdot \cdot \cdot \cdot \cdot}$ in such 
a class of F-theory compactification \cite{TW-06, Berenstein}.

An idea of Ref. \cite{TW-06} is to use Heterotic compactification, and to translate and 
generalise it in the language of F-theory compactification. To be more explicit, imagine a 
Heterotic string compactification on an elliptic fibred Calabi--Yau threefold $(Z, S, \pi)$, 
where $\pi: Z \longrightarrow S$, with a vector bundle $V_3 \oplus V_2$ whose structure group is 
$\SU(3) \times \SU(2) \subset E_8$. $V_3$ and $V_2$ are given by Fourier--Mukai transform of 
spectral data $(C_3, {\cal N}_3)$ and $(C_2, {\cal N}_2)$, where $C_3$ and $C_2$ are divisors of 
$Z$ that are 3-fold and 2-fold covering over $S$, respectively, and ${\cal N}_3$ and ${\cal N}_2$ 
are line bundles on $C_3$ and $C_2$, respectively. For generic complex structure of $Z$, the 
spectral surfaces $C_3$ and $C_2$ are given by 
\begin{equation}
  c_0 + c_2 x + c_3 y = 0, \qquad d_0 + d_2 x = 0,  
\end{equation}
respectively, where 
\begin{equation}
c_k \in \Gamma(S; {\cal O}_S( k K_S + \eta_3)), \qquad 
d_k \in \Gamma(S; {\cal O}_S( k K_S + \eta_2))
\end{equation}
for some divisors $\eta_{2,3}$ of $S$. The F-theory dual of this 
compactification should be given by $(Y_4, B_3)$, where the base 
threefold 
$B_3 = \P \left[ {\cal O}_S(6K_S + \eta_3 + \eta_2) \oplus {\cal O}_S\right]$
is a $\P^1$-fibration over $S$, and the elliptic fibre of 
$\pi: Y_4 \longrightarrow B_3$ is given by \cite{KMV-BM, DW-1, Hayashi-Aspects, Hayashi-Codim}
\begin{equation}
 y^2 = x^3 + f_0 x s^4 + g_0 s^6 + 
     (c_0 s^3 + c_2 s x + c_3 y)(d_0 s^2 + d_2 x), 
\end{equation}
where $s$ is an inhomogeneous coordinate of the $\P^1$-fibre in $B_3 \longrightarrow S$. Now, we 
generalise it to general $B_3$ and its effective divisor $S$, and define 
$\pi: Y_4 \longrightarrow B_3$ by the same equation as above; the coefficients $f_0$, $g_0$, $c_k$'s 
and $d_k$'s, however, are promoted to holomorphic sections on $B_3$ as follows:
\begin{eqnarray}
  c_k \in \Gamma(B_3; {\cal O}_B((k-4)K_B+a+(k-3)S)) , & \quad &
     f_0 \in \Gamma(B_3; {\cal O}_B(-4 K_B - 4S), \\
  d_k \in \Gamma(B_3; {\cal O}_B(-a + (k-2)(K_B+S))) , & \quad & 
     g_0 \in \Gamma(B_3; {\cal O}_B(-6 K_B - 6S),  
\end{eqnarray}
where $a$ is some divisor on $B_3$; this is a generalisation, in that the translation from 
Heterotic string compactification is reproduced by setting $(2K_S + \eta_2) = - a|_S$ and 
$(6K_S +\eta_2 + \eta_3) = S|_S = c_1(N_{S|B_3})$. 

One can read out from the discriminant and singularity of this generalised 
Weierstrass form that there are three distinct matter curves,\footnote{
The linearised analysis \cite{DW-spect} is able to determine 
the defining equation of the spectral cover for associated bundles 
such as $(V_3 \otimes V_2)$ approximately. All the terms except those involving 
$f_0$ or $g_0$ in the defining equation of $\Sigma_{({\bf 6})}$ can be obtained 
in that way.}  
\begin{eqnarray}
 \Sigma_{(\wedge^3 {\bf 6})}: & & d_2|_S = 0, \\
 \Sigma_{(\wedge^2 {\bf 6})}: & & c_3|_S = 0, \\
 \Sigma_{({\bf 6})}: & & 
   (c_3^2 d_0^3 + c_2^2 d_0^2 d_2 - 2 c_0 c_2 d_0 d_2^2 + c_0^2 d_2^3
 + c_3^2 d_0 d_2^2 f_0 - c_3^2 d_2^3 g_0)|_S = 0. 
\end{eqnarray}
Those three curves intersect at $c_3|_S = d_2|_S = 0$ points in $S$.
We can choose the gauge group of the Katz--Vafa type field theory 
(field theory local model) around these matter curves to be 
$E_6$, $D_6$, $A_6$; physics around a $c_3|_S = d_2|_S = 0$ point is described by 
an $E_7$ gauge theory; a non-trivial Higgs bundle background with the structure group 
$\SU(2) \times \U(1) \subset E_7$ breaks the $E_7$ symmetry down to SU(6); Yukawa coupling 
$\Delta W = {\bf 6} \cdot \wedge^2 {\bf 6} \cdot \wedge^3 {\bf 6}$ is generated at each one 
of those $c_3|_S = d_2|_S = 0$  points. 

Such an SU(6) 7-brane in F-theory can be used for SU(5) unification by turning on a line bundle on $S$, so that 
the symmetry is reduced to SU(5); further breaking to the Standard Model gauge group is not impossible, although 
we stay away from such details. There are two possible particle identifications. The first possibility is to 
identify SU(5)-${\bf 10}$ matter fields with the $\wedge^3 {\bf 6}$ representation of SU(6), and $H({\bf 5})$ 
within ${\bf adj}.$ of SU(6) \cite{TW-06}; the other possibility is to find the ${\bf 10}$ matter field  
in $\wedge^2 {\bf 6}$ of SU(6), when the $H({\bf 5})$ field also has to come from the same $\wedge^2{\bf 6}$ 
representation of SU(6); the latter possibility was overlooked in \cite{TW-06}. In any one of those two 
possibilities, Yukawa couplings are generated along the entire matter curve
($\Sigma_{(\wedge^3 {\bf 6})}$ or $\Sigma_{(\wedge^2 {\bf 6})}$), not only at isolated points in the 7-brane $S$ 
(cf \cite{BHV-1}). This makes it impossible to exploit the approximately codimension-1 nature of Yukawa 
matrices from isolated Yukawa points \cite{Hayashi-Codim, HV-flavor, Cecotti}. The idea of \cite{Gaussian} 
(or something similar to the one in \cite{Hayashi-flavor}) may still be implemented in the latter identification
with a tuning $j(\Sigma_{(\wedge^2 {\bf 6})}) \gg 1$, it is desirable to have a separate study.  

\vspace{1cm}

Before closing this section, we compute $h^{3,1}$ for Calabi--Yau fourfolds with such an SU(6) 
unification. We choose $(B_3, [S])$ to be $B_3 = \P^1 \times \P^2$ and $[S] = {\rm pt} \times \P^2$, 
so the result can be compared with $h^{3,1}$ for other classes of compactifications with a rank-5 
symmetry (SO(10) and $\SU(5) \times \U(1)$) in the main text.
The choice of $(B_3, [S])$ above introduces a constraint\footnote{Intuitively, 
this constraint means that the instanton number is distributed equally 
into the hidden and visible sectors; 
$(6K_S + \eta_0) = 0 = - (6K_S + \eta_\infty)$. }
$(3K_S + \eta_2) + (3K_S + \eta_3) = 0$.
\begin{table}[tbp]
\begin{center}
\begin{tabular}{|c|ccccc|}
\hline
$(3K_{\P^2} + \eta_2)$ & $-2$ & $-1$ & 0 & 1 & 2 \\ 
$(3K_{\P^2} + \eta_3)$ & 2 & 1 & 0 & $-1$ & $-2$ \\
\hline
$h^{3,1}$ & 1918 & 1909 & 1905 & 1906 & 1912 \\
\hline
\end{tabular}
\caption{\label{tab:h31-SU(6)}
$h^{3,1}$ of Calabi--Yau fourfolds for SU(6) unification, 
provided $B_3 = \P^1 \times \P^2$ and $[S] = {\rm pt} \times \P^2$. }
\end{center}
\end{table}
See Table~\ref{tab:h31-SU(6)} for the results.

\section{Monodromy around the U(1)-enhancement Limit}
\label{sec:monodromy}

\subsection{$6g-3$ Topological Four-cycles}
\label{ssec:6g-3}

This appendix \ref{sec:monodromy} begins with a brief review. We came to be interested in 
section \ref{ssec:approx-U1-general} in a compact Calabi--Yau fourfold $Y_4$ with its complex 
structure parameter in ${\cal M}_*$ close to the ${\cal M}_*^{\U(1)}$ locus; $Y_4$ contains a local 
geometry of deformed conifold along a curve $\Sigma$, and this local geometry of $Y_4$ is modelled 
by a geometry $Y_{\rm local}$, which is explained shortly. Four-cycles in $Y_{\rm local}$ as well as 
their lift to the global geometry $Y_4$ was studied in \cite{conifold-CY4}; results 
of \cite{conifold-CY4} that we need in section \ref{ssec:approx-U1-general} are reviewed here. 
The review is followed by analysis of monodromy of those cycles and period integrals.  

The local geometry model $Y_{\rm local}$, which is denoted by $\widetilde{X}^\flat$ in \cite{conifold-CY4}, 
is realised as a hypersurface of the total space of a rank-4 vector bundle over a Riemann surface $\Sigma$, 
\begin{equation}
   {\cal L}^{\otimes 3} \oplus {\cal L}^{\otimes 3} \oplus {\cal L}^{\otimes 2} \oplus {\cal L}^{\otimes 4}   
    \longrightarrow \Sigma, 
\label{eq:appC-ambient-space}
\end{equation}
where the Riemann surface $\Sigma$ satisfies $6|(2g(\Sigma)-2)$, and ${\cal L}^{\otimes 6} = K_{\Sigma}$.
The defining equation of $Y_{\rm local}$ in this ambient space is 
\begin{equation}
 Y A_3 = X A_4 + A_6;
\end{equation}
$Y$, $A_3$, $X$ and $A_4$ are the coordinates of the rank-4 fibre of the bundle 
in (\ref{eq:appC-ambient-space}), and 
\begin{equation}
  A_6 \in \Gamma(\Sigma; K_\Sigma) \cong \C^g =: {\cal M}_*^{\rm local}
\end{equation}
governs the complex structure of this local geometry; this $A_6 \in \Gamma(\Sigma; K_{\Sigma})$ descends 
from $A_6 \in \Gamma(B_3; {\cal O}_B(-6K_B))$ on the compact set-up by simple restriction on 
$\Sigma \subset B_3$. 
${\cal L}^{\otimes 6} := {\cal O}_\Sigma(-6K_B|_{\Sigma})$ is the same as $K_\Sigma$, because of 
the adjunction formula for $\Sigma := \left\{A_3 = A_4 = 0 \right\} \subset B_3$. 
The $g$-dimensional space ${\cal M}_*^{\rm local}$ is regarded as the $g$ directions normal to 
${\cal M}_*^{\U(1)}$ in ${\cal M}_*$, at least when $B_3$ is a Fano variety.

Reference \cite{conifold-CY4} identified $6g-3$ four-cycles in this local fourfold geometry 
$Y_{\rm local}$. Let $Z : = \partial \overline{Y_{\rm local}}$ be the boundary, which is a seven 
dimensional manifold over $\R$. Using a long exact sequence 
\begin{equation}
 0 \longrightarrow H_4(Z; \Z)  \longrightarrow H_4(Y_{\rm local}; \Z) \longrightarrow 
  H_4^{BM}(Y_{\rm local}; \Z) \longrightarrow H_3(Z; \Z) \longrightarrow 0, 
\end{equation}
it turns out that both $H_4(Y_{\rm local};\Q)$ and $H_4^{BM}(Y_{\rm local}; \Q)$ are of dimension $4g-3$;  
kernels and cokernels of the homomorphisms in the exact sequence above introduces a filtration 
structure 
\begin{align}
   H_4(Y_{\rm local}; \Q) & \supset \left(  H_4(Y_{\rm local}; \Q) \right)^0 =: 
       {\rm Span}_\Q \left\{ \widetilde{A}_{i=1,\cdots,g}, \widetilde{A}^{'j=1,\cdots,g} \right\}, \\
  H_4(Y_{\rm local}; \Q)  \; / \; \left(  H_4(Y_{\rm local}; \Q) \right)^0  &=: 
   {\rm Span}_\Q \left\{ [\widetilde{B}_\ell] \; | \; \ell = 1, \cdots, 2g-3 \right\},  
\end{align}
and 
\begin{align}
   H_4^{BM}(Y_{\rm local}; \Q) & \supset \left(  H_4^{BM}(Y_{\rm local} ; \Q) \right)^0 =: 
       {\rm Span}_\Q \left\{ \widetilde{B}'_\ell \; | \;   \ell=1,\cdots,2g-3 \right\}, \\
   H_4^{BM}(Y_{\rm local}; \Q) \; / \; \left(  H_4^{BM}(Y_{\rm local} ; \Q) \right)^0 &=: 
       {\rm Span}_\Q \left\{ [\widetilde{C}]^{i=1,\cdots,g}, \; [\widetilde{C}']_{j=1,\cdots, g} \right\};   
\end{align}
Overall, $2g + (2g-3) + 2g$ four-cycles, $\widetilde{A}$'s, $\widetilde{B}$'s and $\widetilde{C}$'s 
are identified in either $H_4(Y_{\rm local})$ or $H_4^{BM}(Y_{\rm local})$.
The intersection pairing $H_4(Y_{\rm local}; \Q) \times H_4^{BM}(Y_{\rm local}; \Q) \longrightarrow \Q$ vanishes 
on $\left( H_4(Y_{\rm local}; \Q) \right)^0 \times \left( H_4^{BM}(Y_{\rm local}; \Q) \right)^0$.

The four-cycles $\tilde{A}_i$'s and $\widetilde{A}^{'j}$'s are the nearly vanishing $S_3$ cycle (often 
referred to as the $A$-cycle) of deformed conifold fibred over the one-cycles $\alpha_i$'s and $\beta^j$ 
of the genus $g$ curve $\Sigma$. Four-cycles $\widetilde{B}_\ell$'s ($\ell=1,\cdots, 2g-3$), on the other 
hand, are topologically $S_4$, and arise in the form of $S_3$ fibred over intervals $I_{\ell}$ on $\Sigma$;
the interval $I_{\ell}$ ($\ell=1,\cdots, 2g-3$) is stretched between a pair of points 
$p_0, p_\ell \in \Sigma$, where $\left\{p_0, p_{\ell=1,\cdots, 2g-3} \right\} \subset \Sigma$ are the zeros 
of the section $A_6 \in \Gamma(\Sigma; K_\Sigma)$. Choice of the interval $I_\ell$ (between $p_0$ and $p_\ell$)
comes with freedom of $+ H_1(\Sigma; \Z)$; this is how the filtration structure arises in 
$H_4(Y_{\rm local}; \Z)$; by choosing an interval $I_\ell$, a representative four-cycle $\widetilde{B}_\ell$ 
is chosen for a quotient class $[\widetilde{B}_\ell]$. Geometric description of the four-cycles 
$\widetilde{C}$'s is given later in this appendix. 

Period integral is defined for all of these $6g-3$ four-cycles in $Y_{\rm local}$; their period integrals
should depend on the $g$ independent moduli of ${\cal M}_*^{\rm local}$. Let $\{ \lambda^i \}_{i=1,\cdots, g}$ be the 1-forms of $\Sigma$ normalised  
so that $\int_{\alpha_i} \lambda^j = \delta_i^{\; j}$. By parameterising $A_6$ on $\Sigma$ (and parameterising 
also ${\cal M}_*^{\rm local}$) as 
\begin{equation}
  A_6 = \sum_i z_i \lambda^i, 
  \label{eq:appC-def-coordinate-zi}
\end{equation}
we can write down the period integrals for the four-cycles $\widetilde{A}$'s  as 
\begin{equation}
 \Pi_{\widetilde{A}_i} = z_i, \qquad \qquad \Pi_{\widetilde{A}^{'j}} = \tau^{jk} z_k,
\end{equation}
where $\tau^{jk} := \int_{\beta^j} \lambda^k$ is the period matrix of the curve $\Sigma$.
The period integrals for the four-cycles $\widetilde{B}_\ell$'s are given by 
\begin{equation}
  \Pi_{\widetilde{B}_\ell} = \tilde{\mu}(p_\ell)^i z_i,
\end{equation}
using the lift of Abel--Jacobi map
\begin{equation}
  \tilde{\mu}: \Sigma \ni q \longmapsto 
  \left( \int_{p_0}^q \lambda^1, \int_{p_0}^q \lambda^2, \cdots, \int_{p_0}^q \lambda^g \right) \in \C^g;  
\end{equation}
although the Abel--Jacobi map itself depends only on the parameters in ${\cal M}_*^{\U(1)}$, 
dependence on the ${\cal M}_*^{\rm local}$ parameters in $\Pi_{\widetilde{B}_\ell}$ comes in through 
$p_\ell$'s as well as $z_i$'s.
When the interval $I_\ell$ from $p_0$ to $p_\ell$ is changed by $H_1(\Sigma; \Z)$, the period integral 
$\Pi_{\widetilde{B}_\ell}$ changes by $n^i \Pi_{\widetilde{A}_i} + m_j \Pi_{\widetilde{A}^{'j}}$ for some 
$n^i, m_j \in \Z$. This transformation constitutes a part of the modular group \cite{conifold-CY4}.

Some of the four-cycles $\widetilde{A}$'s in $H_4(Y_{\rm local})$ in the local geometry may not be 
regarded as topological cycles $H_4(Y_4)$ in the global geometry; all the $\widetilde{A}$'s
can be deformed to be topological cycles in $Z = \partial \overline{Y_{\rm local}}$, and such a 
four-cycle may, in principle, be obtained as a boundary of a five-cycle in $Y_4 \backslash Y_{\rm local}$.
The relation $(-\Delta h^{3,1}) = g$ for the global geometry $Y_4$ (which holds at least when $B_3$ is 
a Fano), and its consequence $\tilde{h}^{2,1} = 0$, in particular, implies that all of the $2g$ four-cycles 
$\widetilde{A}$'s remain to be topological four-cycles of the global geometry $Y_4$. Similarly, 
the four-cycles $\widetilde{C}$'s in $H_4^{BM}(Y_{\rm local})$ can be regarded as topological cycles 
of $Y_4$, only when their boundaries in $H_3(\partial \overline{Y_{\rm local}})$ are obtained also as 
boundaries of some cycles in $H_4^{BM}(Y_4 \backslash \overline{Y_{\rm local}})$. The Poincare duality 
indicates, however, that all of these $2g$ four-cycles $\widetilde{C}$'s are also lifted to those 
in the global geometry $Y_4$,
at least when $B_3$ is Fano. In the conifold transition at the $A_6 \longrightarrow 0$ limit, those
$2g + (2g-3) + 2g$ topological four-cycles shrink, and one four-cycle ($\P^1$ for small resolution over 
the curve $\Sigma$) emerges in the global geometry $Y_4$; $-\Delta h^{3,1} = -\Delta h^{1,3} = g$, and 
$-\Delta h^{2,2} = (4g-3) -1 = 4 (-\Delta h^{3,1} - \Delta h^{1,1})$. See \cite{conifold-CY4} for more 
information.  

\subsection{Monodromy}
\label{ssec:cycle-monodromy}

In order to study monodromy of those four-cycles in the local geometry $Y_{\rm local}$, we assume that 
$\Sigma$ is a hyperelliptic curve in this appendix \ref{ssec:cycle-monodromy}: 
\begin{equation}
  t^2 = P(s); \qquad P(s) = - \prod_{i=1}^{2g+2}(s-s_i);
\end{equation}
we further assume that all the $s_i$'s are real valued, and 
\begin{equation}
0 < s_1 <s_2 \ll s_3 < s_4 \ll \cdots \ll s_{2g-1} < s_{2g} \ll s_{2g+1} < s_{2g+2}.
\end{equation}
although a higher genus curve $\Sigma$ is not always in the form of a hyperelliptic curve, complex 
structure of $\Sigma$ can be continuously deformed from the one chosen above; since we are interested 
primarily in questions of topological nature, it is enough to study for $\Sigma$ given above.
\begin{figure}[tbp]
\begin{center}
  \includegraphics[width=.8\linewidth]{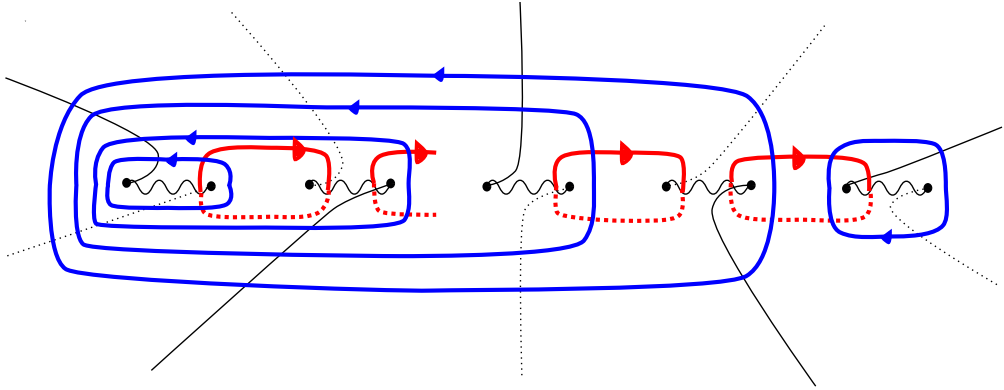}  
\caption{\label{fig:hyperelliptic} (colour online) A hyperelliptic curve $\Sigma$ is obtained by 
gluing together two sheets of this $s$-plane ($\C \cup \{ \infty \}$) along the branch cuts (wavy 
lines) between $s_1$--$s_2$, $s_3$--$s_4$, $\cdots$, $s_{2g+1}$--$s_{2g+2}$; this picture is drawn for 
the case with $g=4$. Description of line bundles ${\cal L}^{\otimes k}$ requires branch cuts in the 
$U^{(s)}$ patch; the cuts for this purpose are drawn in thin (solid or dotted) lines in this figure. 
Thick grey (red) loops, from left to right, are the cycles $\alpha_1$, $\alpha_2$ (drawn 
partially), $\alpha_{g-1}$ and $\alpha_g$.
Thick dark (blue) loops, from left to right, are $\beta^1$, $\beta^2$, $\beta^{g-1}$ and $\beta^g$; 
one more remaining loop in the thick dark (blue) line at the right end of this picture is 
$\beta^{'g}$; $\beta^g$ and $\beta^{'g}$ are isomorphic in $\Sigma$, but they are not within the 
$U^{(s)}$ patch.
}
\end{center}
\end{figure}

Before getting into the study of monodromy, we first need to have a concrete construction of 
the cycles $\widetilde{C}^i$'s ($i=1,\cdots, g$), whose monodromy we are interested in. 
Math preparation is thus in order here. The line bundle $K_{\Sigma} = {\cal L}^{\otimes 6}$ can be 
described by three Zariski open patches of $\Sigma$.
\begin{eqnarray}
   U^{(s)}: \quad  t \neq 0, \infty; \qquad 
&  U^{(t)}: \quad   P'(s) \neq 0, \; s \neq \infty; \qquad 
&  U^{(\infty)}: \quad s \neq 0, s_i.  
\end{eqnarray}
Sections of $K_\Sigma$ are written down in the form of $A = a^{(s)} ds$, $A^{(t)}dt$ and 
$A^{(\infty)} d(1/s)$ in the $U^{(s)}$, $U^{(t)}$ and $U^{(\infty)}$ patch, respectively; 
these trivialisation descriptions are identified by using transition functions:
\begin{equation}
  a^{(t)} = \frac{2t}{P'(s)} a^{(s)}, \qquad 
  a^{(\infty)} = (-s^2) a^{(s)}.
\label{eq:appC-KSigma-trans-fcn}
\end{equation}
$H^0(\Sigma; K_\Sigma)$ is of $g$ dimensions, and are of the form\footnote{
There must be a linear relation between $\{ c_0, c_1, \cdots, c_{g-1} \}$ and 
$z_i$'s ($i=1,\cdots, g$) in (\ref{eq:appC-def-coordinate-zi}), but we do not need to know it 
in detail.}   
\begin{equation}
 A_6^{(s)}ds  = \frac{c_0 + c_1 s + \cdots + c_{g-1} s^{g-1} }{t} ds 
  = \frac{c_0 (1/s)^{g-1} + \cdots c_{g-1}}{(t/s^{g+1})} d(1/s) = A_6^{(\infty)} d(1/s).
\label{eq:appC-hol-A6-parametrize}
\end{equation}

The fibre coordinates $Y$ of ${\cal L}^{\otimes 3}$ and $X$ of ${\cal L}^{\otimes 2}$, for example, 
become $Y^{(s)}$, $Y^{(t)}$ and $Y^{(\infty)}$, and $X^{(s)}$, $X^{(t)}$ and $X^{(\infty)}$, respectively, 
in the trivialisation patches, and are identified between the overlapping patches as in 
\begin{equation}
  Y^{(t)} = \left(\frac{2t}{P'(s)}\right)^{3/6} Y^{(s)}, \qquad 
  X^{(t)} = \left(\frac{2t}{P'(s)} \right)^{2/6} X^{(s)}.
\label{eq:trans-fcn-fract-bndl}
\end{equation}
Branch cuts are introduced in these $U^{(s)}$, $U^{(t)}$ and $U^{(\infty)}$; 
see Figure~\ref{fig:hyperelliptic} for the branch cuts in the $U^{(s)}$ patch; the coordinates 
$Y^{(s)}$ and $X^{(s)}$ at one point in $U^{(s)} \subset \Sigma$ and the coordinates $Y^{(s)'}$ and 
$X^{(s)'}$ at the same point that we reach after circling around a branch point (where $t = 0$) 
by phase $+2\pi$ are identified through
\begin{equation}
  Y^{(s)'} = Y^{(s)} \times \zeta_6^{-3}, \qquad X^{(s)'} = X^{(s)} \times \zeta_6^{-2}; 
   \qquad \zeta_6 := e^{\frac{2\pi i}{6}}.
\end{equation}
Fibre coordinates need to be identified through similar relations also across the branch cuts in 
$U^{(t)}$ and $U^{(\infty)}$. Equations (\ref{eq:trans-fcn-fract-bndl}) are made well-defined in this 
way. The same holds true also for the fibre coordinates $A_3$ and $A_4$.

Let us take a point $q \in U^{(s)} \subset \Sigma$. The local fourfold geometry $Y_{\rm local}$  
has a three-dimensional fibre 
\begin{equation}
 Y^{(s)} A_3^{(s)} = X^{(s)} A_4^{(s)} + A_6^{(s)};
\label{eq:defrm-cnfd-trivialpatch}
\end{equation}
$Y^{(s)}$, $A_3^{(s)}$, $X^{(s)}$ and $A_4^{(s)}$ are coordinates, while $A_6^{(s)}$ is a parameter. 
This is a deformed conifold, and there is a canonical choice of compact three-cycle and 
a semi-canonical choice of non-compact three-cycle intersecting at one point; they are referred 
to as $A$-cycle and $B$-cycle; the choice of the $B$-cycle is not canonical, in that 
the $B$-cycle is deformed to be $B+A$ topologically, when the complex phase of the parameter 
$A_6^{(s)}$ changes as $A_6^{(s)} \longrightarrow A_6^{(s)} \times e^{i \alpha}$, $\alpha \in [0, 2\pi]$, 
as well-known in deformed conifold. 

Now, we are ready to provide description of the $2g$ remaining four-cycles, 
$\widetilde{C}^{i=1,\cdots,g}$ and $\widetilde{C}^{'}_{j=1,\cdots,g}$ in $Y_{\rm local}$. Since this task 
is topological in nature, we can choose the parameter $A_6 \in {\cal M}_*^{\rm local}$ 
arbitrarily; we choose it to be 
\begin{equation}
  A_6 = A_6^{*} := \epsilon \frac{(s-s_*)^{g-1}}{t} ds, \qquad \qquad s_{2g+2} \ll s_* \in \R
\end{equation}
for now. First of all, the cycles $\widetilde{A}_{i=1,\cdots,g}$ and 
$\widetilde{A}^{'j=1,\cdots,g}$ are the $A$-cycle fibred over the one-cycles $\alpha_{i=1,\cdots,g}$
and $\beta^{j=1,\cdots,g}$ in $\Sigma$, as in the appendix \ref{ssec:6g-3}; see 
Figure~\ref{fig:hyperelliptic} for how to choose the basis of $H_1(\Sigma; \Z)$.
\begin{figure}[tbp]
\begin{center}
\begin{tabular}{ccc}
    \includegraphics[width=.4\linewidth]{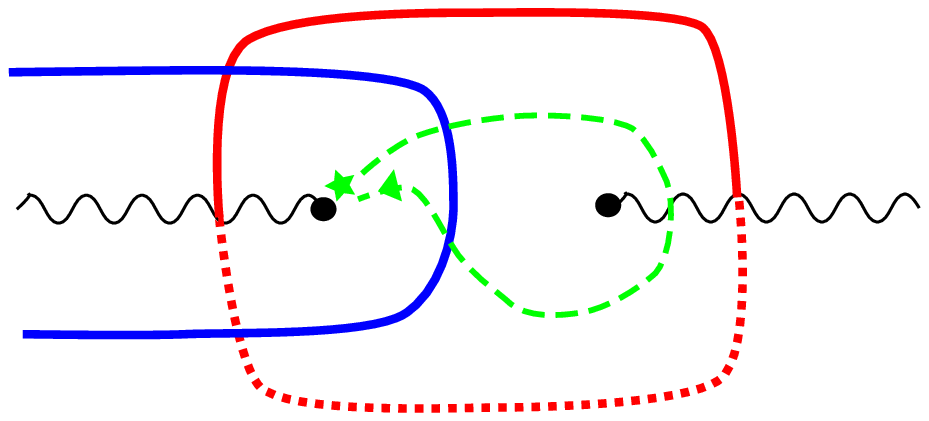}  & & 
  \includegraphics[width=.4\linewidth]{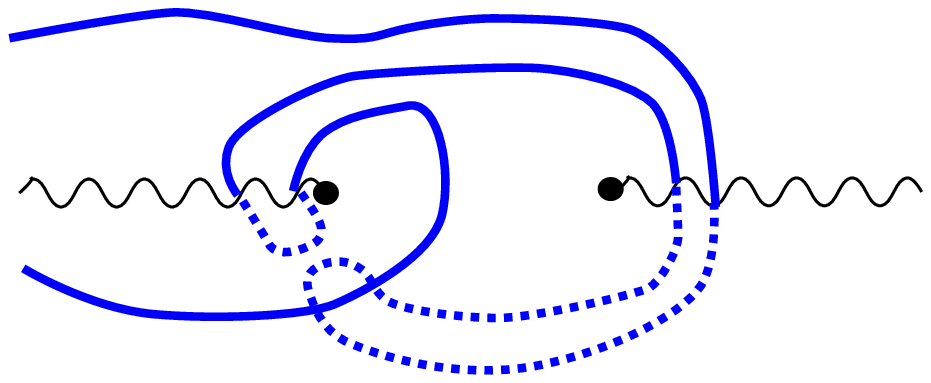}  \\
 (a) & & (b)
\end{tabular}
\caption{\label{fig:loop-monodromy} (colour online) 
(a) A loop $\gamma_k$ ($k=1,\cdots, g-1$) in the ${\cal M}_*^{\rm local} \cong \C^g$ parameter space 
is given by continuously changing the parameter $a_k$ in the $s$-plane, starting from $s=s_{2k}$, 
going around $s=s_{2k+1}$ and returning to $s=s_{2k}$, as shown by a dashed (green) arrow line in 
this picture. At the end of this deformation procedure, the loop $\beta^k$ has been deformed in the 
way shown in (b) by a thick solid (blue) line. }
\end{center}
\end{figure}
Secondly, for this choice of $A_6^{*}$, the cycles $\widetilde{B}_{\ell=1,\cdots, 2g-3}$ are 
all located in the $s \sim s_* \gg s_{2g+2}$ region; we see after constructing $\widetilde{C}$'s 
that $\widetilde{B}_{\ell}$'s and $\widetilde{A}$--$\widetilde{C}$'s are mutually orthogonal 
in the intersection form. Finally, we claim that the $B$-cycle of the deformed conifold comes 
back to itself, not to $B+mA$ with $m \neq 0$, after a point $q \in \Sigma$ moves along any one 
of $\{ \alpha_i, \beta^j \}$. To verify this claim, first note that the loop $\beta^k$ 
crosses the branch cut in $U^{(s)}$ for $k$-times in the counter-clockwise direction; this means 
that the fibre coordinate of ${\cal L}^{\otimes a}$ at the end of a loop along $\beta^k$ is 
$\zeta_6^{-a \times k} = e^{- 2\pi i \times \frac{a \cdot k}{6}}$ times the one at the beginning of the loop. 
The parameter $(A_6^{*})^{(s)}$, on the other hand, changes its phase by $e^{- 2\pi i k}$ due to the 
factor $1/t$. Those two phases on both sides of (\ref{eq:defrm-cnfd-trivialpatch}) cancel, and 
there is no net change in the phase of the deformation parameter (not even a multiple of $2\pi$) 
along the loop $\beta^k$. Thus, the $B$-cycle comes back to itself. The $B$-cycle fibred over 
$\beta^k$ ($k=1,\cdots, g$) forms a four-cycle $\widetilde{C}^k$.  Similarly, we note that 
the loop $\alpha_k$ crosse the branch cuts in $U^{(s)}$ for $(-1)$ times in the counter-clockwise 
direction. The parameter $(A_6^*)^{(s)}$ changes its phase by $e^{+2\pi i}$ due to the factor $1/t$, 
on the other hand. Those two effects cancel, and there is no net change in the phase. Thus, 
the $B$-cycle comes back to itself at the end of the loop $\alpha_k$. This is how a four-cycle 
$\widetilde{C}^{'}_{k}$ is obtained ($k=1,\cdots, g$). By construction, 
$\widetilde{A}_i \cdot \widetilde{C}^j = \delta_i^{\; j}$, 
$\widetilde{A}^{'k} \cdot \widetilde{C}'_h = \delta^k_{\; h}$, and the intersection number vanishes 
for all other combinations of $\widetilde{A}$'s and $\widetilde{C}$'s. 

Finally, we study monodromy of those four-cycles in $Y_{\rm local}$. Monodromy is studied for loops 
departing and returning to a reference point in ${\cal M}_*^{\rm local} = \C^g$, and we choose 
\begin{equation}
 A_6 = A_6^{**} := \epsilon \frac{ (s-s_2)(s-s_4) \cdots (s-s_{2(g-1)}) }{t} ds 
\end{equation}
as the reference point.\footnote{The point $A_6^{*}$ in ${\cal M}_*^{\rm local}$ is useful in that 
all the $6g-3$ four-cycles can be constructed systematically. The point $A_6^{**}$ is more convenient 
as the reference point of the monodromy study. This is just a matter of convenience. }\raisebox{5pt}{,}\footnote{
The degree $2g-2$ divisor corresponding to this choice of $A_6=A_6^{**}$ is a collection of the $g-1$ 
points $\{ (s,t) = (s_{2i},0) \; | \; i=1,\cdots, g-1 \} \subset \Sigma$ with multiplicity 2 for all of 
them. They are the $2g-2$ points $\{p_0, \cdots, p_{2g-3}\}$ used in construction of the four-cycles 
$\widetilde{B}_{\ell}$'s \cite{conifold-CY4}.} At this reference point, 
let $\widetilde{A}_k$ and $\widetilde{A}^{'k}$ ($k=1,\cdots, g$) be the four-cycle given by the $A$-cycle 
along $\alpha_k$ and $\beta^k$, respectively. $g-1$ more four-cycles $\widetilde{C}^k$ ($k=1,\cdots, g-1$) 
are the $B$-cycle fibred over $\beta^k + k \alpha_k$ in $\Sigma$; this loop in $\Sigma$ crosses $+k$ times 
along $\beta^k$ and $k \times (-1)$ times along $k \alpha_k$, and there is no net change in the phase of 
$A_6^{**}$ along the loop. One more four-cycle, $\widetilde{C}^g$, is the $B$-cycle fibred over a one-cycle 
$\beta^{'g} \sim \beta^g$ on $\Sigma$ shown in Figure~\ref{fig:hyperelliptic}. We will focus on monodromy 
associated with those $3g$ four-cycles. 

It is convenient to adopt the following parameterisation of ${\cal M}_*^{\rm local} \cong 
H^0(\Sigma; K_\Sigma)$:
\begin{equation}
 A_6 = \epsilon \frac{(s-a_1)(s-a_2) \cdots (s-a_{g-1})}{t} ds, \qquad 
   \left\{ (\epsilon, a_1, \cdots, a_{g-1}) \right\}  \in \C^g.
\end{equation}
The reference point $A_6^{**}$ corresponds to choosing $a_k = s_{2k}$ for $k=1,\cdots, g-1$.
Loops $\gamma_k$ for $k=1,\cdots, g-1$ in ${\cal M}_*^{\rm local}$ are such that $a_k$ is changed 
continuously in the $s$-plane in the way designated in Figure~\ref{fig:loop-monodromy}~(a), while 
the value of $\epsilon$ and all other $a_m$'s ($m = 1,\cdots, g-1$ but $m \neq k$) are held fixed. 
One can keep track of topology of the four-cycles $\widetilde{A}$'s and $\widetilde{C}$'s along the loop 
$\gamma_k$ in ${\cal M}_*^{\rm local}$, by deforming the one-cycles 
$\beta^m + m \alpha_m$ ($m=1,\cdots, g-1$) and $\beta^{'g}$ so that the zero of $A_6^{(s)}$ is avoided. 
This is enough to conclude that all the $3g$ four-cycles, namely, $\widetilde{A}$'s, $\widetilde{A}'$'s and 
$\widetilde{C}$'s, remain the same at the end of a loop $\gamma_k$ except $\widetilde{C}^k$. Furthermore, 
because the one-cycle $\beta^k$ needs to be deformed as in Figure~\ref{fig:loop-monodromy}~(b) at the end of 
the loop $\gamma_k$, there is a non-trivial monodromy 
\begin{equation}
 \gamma_k:  \left( \widetilde{A}_k, \widetilde{C}^k \right) \rightarrow 
   \left( \widetilde{A}_k, \widetilde{C}^k \right)
   \left( \begin{array}{cc} 1 & 1 \\ & 1 \end{array} \right), \qquad 
  \left( \widetilde{A}_{m\neq k}, \widetilde{A}^{'j}, \widetilde{C}^{m \neq k} \right) \rightarrow 
  \left( \widetilde{A}_{m \neq k}, \widetilde{A}^{'j}, \widetilde{C}^{m \neq k} \right).
\end{equation}

We study monodromy along one more loop $\gamma_\epsilon$ in ${\cal M}_*^{\rm local}$, which is to change the phase 
of the parameter $\epsilon$ by $2\pi$, while all the $a_m$'s with $m=1,\cdots, g-1$ are held fixed. 
Topological cycles $\widetilde{A}_k$'s for $k=1,\cdots, g$ remain the same under the 
complex structure deformation along $\gamma_\epsilon$. Topological cycles $\widetilde{C}^k$'s are not, 
however. These cycles are all in the form of the $B$-cycle fibred over some one-cycle in $\Sigma$; after 
complex structure deformation along $\gamma_\epsilon$, the original $B$-cycle comes back as $B+A$-cycle. 
This means that 
\begin{equation}
 \gamma_\epsilon: \widetilde{C}^{k \neq g} \longmapsto \widetilde{C}^{k}
     + k \widetilde{A}_k + \widetilde{A}^{'k}, \qquad 
  \widetilde{C}^g \longmapsto \widetilde{C}^g + \widetilde{A}^{'g}.
\end{equation}
This is enough to conclude that the period integrals depend on $\epsilon$ as 
\begin{equation}
 \Pi_{\widetilde{A}_k}, \quad \Pi_{\widetilde{A}^{'k}} \sim \epsilon, \qquad 
 \Pi_{\widetilde{C}^{k \neq g}} \sim \left( \tau^{km} z_m + k z_k \right) \ln (\epsilon), 
 \qquad \Pi_{\widetilde{C}^g} \sim \left( \tau^{gm} z_m  \right)\ln (\epsilon).
\end{equation}

It is a much more involved problem to determine the full monodromy group repreesnted on the space of $6g-3$ 
four-cycles, and also the period integrals. We do not do so in this article, since we do not need such a 
thorough analysis for the sketchy argument in the main text.  

\end{document}